\documentclass[longauth]{aa}

\usepackage{graphicx}
\usepackage{natbib}
\usepackage{scalerel}
\usepackage[table]{xcolor}

\usepackage{txfonts}
\usepackage[pdfencoding=auto,psdextra]{hyperref}
\hypersetup{
    colorlinks=true,
    linkcolor=blue,
    filecolor=magenta,      
    urlcolor=blue,
    citecolor=blue
} 

\usepackage{url}
\usepackage{amsmath}
\usepackage{orcidlink}
\makeatletter
\renewcommand*\aa@pageof{, page \thepage{} of \pageref*{LastPage}}
\makeatother

\newcommand{\software}[1]{\texttt{#1}}

\usepackage[utf8]{inputenc}

\usepackage{euclid}
\let\linenumbers\relax
\defcitealias{Sanchez-Alarcon23}{SA23b}
\defcitealias{Sanchez-Alarcon25}{SA25}
\defcitealias{Watkins19}{W19}

\begin{document}
\nolinenumbers

\title{Euclid Quick Data Release (Q1)}

  \subtitle{Disc breaks highlight the structural evolution of galaxies
through time}

%%%% Version Wednesday 22nd of July 2026 05:47:11 PM UT
%%%% Assumes the new A&A style file from Oct 2025 or later
%%%% Please do not edit the author list -- contact ECEB Bureau for changes
\newcommand{\orcid}[1]{} %% if already defined in aa.cls: comment, or use renewcommand			   
\author{Euclid Collaboration: P.~M.~Sanchez-Alarcon\orcid{0000-0002-6278-9233}\thanks{pablo.m.sanchezalarcon@nasa.gov}\inst{\ref{aff1},\ref{aff2}, \ref{aff11}}
\and J.~Rom\'an\orcid{0000-0002-3849-3467}\inst{\ref{aff3}}
\and S.~Comeron\orcid{0000-0002-7398-4907}\inst{\ref{aff2},\ref{aff1}}
\and J.~H.~Knapen\orcid{0000-0003-1643-0024}\inst{\ref{aff1},\ref{aff2}}
\and M.~Huertas-Company\orcid{0000-0002-1416-8483}\inst{\ref{aff1},\ref{aff2},\ref{aff5},\ref{aff6}}
\and F.~Buitrago\orcid{0000-0002-2861-9812}\inst{\ref{aff7},\ref{aff8},\ref{aff9}}
\and M.~Montes\orcid{0000-0001-7847-0393}\inst{\ref{aff10}}
\and A.~S.~Borlaff\orcid{0000-0003-3249-4431}\inst{\ref{aff11},\ref{aff12}}
\and P.~M.~Marcum\inst{\ref{aff11}}
\and J.~Junais\orcid{0000-0002-7016-4532}\inst{\ref{aff13},\ref{aff1},\ref{aff2}}
\and E.~Duran-Camacho\orcid{0000-0002-3153-0536}\inst{\ref{aff1},\ref{aff2}}
\and M.~N.~Le\orcid{0009-0003-0674-9813}\inst{\ref{aff1},\ref{aff2}}
\and M.~Dunn\orcid{0000-0001-5374-1644}\inst{\ref{aff1},\ref{aff2},\ref{aff14}}
\and H.~Dannerbauer\orcid{0000-0001-7147-3575}\inst{\ref{aff1},\ref{aff2}}
\and B.~Altieri\orcid{0000-0003-3936-0284}\inst{\ref{aff15}}
\and A.~Amara\inst{\ref{aff16}}
\and S.~Andreon\orcid{0000-0002-2041-8784}\inst{\ref{aff17}}
\and N.~Auricchio\orcid{0000-0003-4444-8651}\inst{\ref{aff18}}
\and C.~Baccigalupi\orcid{0000-0002-8211-1630}\inst{\ref{aff19},\ref{aff20},\ref{aff21},\ref{aff22}}
\and M.~Baldi\orcid{0000-0003-4145-1943}\inst{\ref{aff23},\ref{aff18},\ref{aff24}}
\and A.~Balestra\orcid{0000-0002-6967-261X}\inst{\ref{aff25}}
\and S.~Bardelli\orcid{0000-0002-8900-0298}\inst{\ref{aff18}}
\and P.~Battaglia\orcid{0000-0002-7337-5909}\inst{\ref{aff18}}
\and A.~Biviano\orcid{0000-0002-0857-0732}\inst{\ref{aff20},\ref{aff19}}
\and M.~Bolzonella\orcid{0000-0003-3278-4607}\inst{\ref{aff18}}
\and E.~Branchini\orcid{0000-0002-0808-6908}\inst{\ref{aff26},\ref{aff27},\ref{aff17}}
\and M.~Brescia\orcid{0000-0001-9506-5680}\inst{\ref{aff28},\ref{aff29}}
\and J.~Brinchmann\orcid{0000-0003-4359-8797}\inst{\ref{aff30},\ref{aff31},\ref{aff32}}
\and S.~Camera\orcid{0000-0003-3399-3574}\inst{\ref{aff33},\ref{aff34},\ref{aff35}}
\and G.~Ca\~nas-Herrera\orcid{0000-0003-2796-2149}\inst{\ref{aff36},\ref{aff37}}
\and V.~Capobianco\orcid{0000-0002-3309-7692}\inst{\ref{aff35}}
\and C.~Carbone\orcid{0000-0003-0125-3563}\inst{\ref{aff38}}
\and J.~Carretero\orcid{0000-0002-3130-0204}\inst{\ref{aff39},\ref{aff40}}
\and S.~Casas\orcid{0000-0002-4751-5138}\inst{\ref{aff41},\ref{aff42}}
\and M.~Castellano\orcid{0000-0001-9875-8263}\inst{\ref{aff43}}
\and G.~Castignani\orcid{0000-0001-6831-0687}\inst{\ref{aff18}}
\and S.~Cavuoti\orcid{0000-0002-3787-4196}\inst{\ref{aff29},\ref{aff44}}
\and K.~C.~Chambers\orcid{0000-0001-6965-7789}\inst{\ref{aff45}}
\and A.~Cimatti\inst{\ref{aff46}}
\and C.~Colodro-Conde\inst{\ref{aff1}}
\and G.~Congedo\orcid{0000-0003-2508-0046}\inst{\ref{aff47}}
\and C.~J.~Conselice\orcid{0000-0003-1949-7638}\inst{\ref{aff48}}
\and L.~Conversi\orcid{0000-0002-6710-8476}\inst{\ref{aff49},\ref{aff15}}
\and Y.~Copin\orcid{0000-0002-5317-7518}\inst{\ref{aff50}}
\and A.~Costille\inst{\ref{aff51}}
\and F.~Courbin\orcid{0000-0003-0758-6510}\inst{\ref{aff52},\ref{aff53}}
\and H.~M.~Courtois\orcid{0000-0003-0509-1776}\inst{\ref{aff54}}
\and M.~Cropper\orcid{0000-0003-4571-9468}\inst{\ref{aff55}}
\and A.~Da~Silva\orcid{0000-0002-6385-1609}\inst{\ref{aff56},\ref{aff57}}
\and H.~Degaudenzi\orcid{0000-0002-5887-6799}\inst{\ref{aff58}}
\and G.~De~Lucia\orcid{0000-0002-6220-9104}\inst{\ref{aff20}}
\and C.~Dolding\orcid{0009-0003-7199-6108}\inst{\ref{aff55}}
\and H.~Dole\orcid{0000-0002-9767-3839}\inst{\ref{aff59}}
\and F.~Dubath\orcid{0000-0002-6533-2810}\inst{\ref{aff58}}
\and C.~A.~J.~Duncan\orcid{0009-0003-3573-0791}\inst{\ref{aff47}}
\and X.~Dupac\inst{\ref{aff15}}
\and S.~Dusini\orcid{0000-0002-1128-0664}\inst{\ref{aff60}}
\and S.~Escoffier\orcid{0000-0002-2847-7498}\inst{\ref{aff61}}
\and M.~Fabricius\orcid{0000-0002-7025-6058}\inst{\ref{aff62},\ref{aff63}}
\and M.~Farina\orcid{0000-0002-3089-7846}\inst{\ref{aff64}}
\and R.~Farinelli\inst{\ref{aff18}}
\and S.~Ferriol\inst{\ref{aff50}}
\and F.~Finelli\orcid{0000-0002-6694-3269}\inst{\ref{aff18},\ref{aff65}}
\and S.~Fotopoulou\orcid{0000-0002-9686-254X}\inst{\ref{aff66}}
\and M.~Frailis\orcid{0000-0002-7400-2135}\inst{\ref{aff20}}
\and E.~Franceschi\orcid{0000-0002-0585-6591}\inst{\ref{aff18}}
\and M.~Fumana\orcid{0000-0001-6787-5950}\inst{\ref{aff38}}
\and S.~Galeotta\orcid{0000-0002-3748-5115}\inst{\ref{aff20}}
\and K.~George\orcid{0000-0002-1734-8455}\inst{\ref{aff67}}
\and B.~Gillis\orcid{0000-0002-4478-1270}\inst{\ref{aff47}}
\and C.~Giocoli\orcid{0000-0002-9590-7961}\inst{\ref{aff18},\ref{aff24}}
\and J.~Gracia-Carpio\inst{\ref{aff62}}
\and A.~Grazian\orcid{0000-0002-5688-0663}\inst{\ref{aff25}}
\and F.~Grupp\inst{\ref{aff62},\ref{aff63}}
\and S.~Gwyn\orcid{0000-0001-8221-8406}\inst{\ref{aff68}}
\and W.~G.~Hartley\inst{\ref{aff58}}
\and S.~V.~H.~Haugan\orcid{0000-0001-9648-7260}\inst{\ref{aff69}}
\and J.~Hoar\inst{\ref{aff15}}
\and H.~Hoekstra\orcid{0000-0002-0641-3231}\inst{\ref{aff37}}
\and W.~Holmes\inst{\ref{aff70}}
\and F.~Hormuth\inst{\ref{aff71}}
\and A.~Hornstrup\orcid{0000-0002-3363-0936}\inst{\ref{aff72},\ref{aff73}}
\and K.~Jahnke\orcid{0000-0003-3804-2137}\inst{\ref{aff74}}
\and M.~Jhabvala\inst{\ref{aff75}}
\and B.~Joachimi\orcid{0000-0001-7494-1303}\inst{\ref{aff76}}
\and E.~Keih\"anen\orcid{0000-0003-1804-7715}\inst{\ref{aff77}}
\and S.~Kermiche\orcid{0000-0002-0302-5735}\inst{\ref{aff61}}
\and A.~Kiessling\orcid{0000-0002-2590-1273}\inst{\ref{aff70}}
\and B.~Kubik\orcid{0009-0006-5823-4880}\inst{\ref{aff50}}
\and M.~K\"ummel\orcid{0000-0003-2791-2117}\inst{\ref{aff63}}
\and M.~Kunz\orcid{0000-0002-3052-7394}\inst{\ref{aff78}}
\and H.~Kurki-Suonio\orcid{0000-0002-4618-3063}\inst{\ref{aff79},\ref{aff80}}
\and A.~M.~C.~Le~Brun\orcid{0000-0002-0936-4594}\inst{\ref{aff81}}
\and S.~Ligori\orcid{0000-0003-4172-4606}\inst{\ref{aff35}}
\and P.~B.~Lilje\orcid{0000-0003-4324-7794}\inst{\ref{aff69}}
\and V.~Lindholm\orcid{0000-0003-2317-5471}\inst{\ref{aff79},\ref{aff80}}
\and I.~Lloro\orcid{0000-0001-5966-1434}\inst{\ref{aff82}}
\and M.~Magliocchetti\orcid{0000-0001-9158-4838}\inst{\ref{aff64}}
\and G.~Mainetti\orcid{0000-0003-2384-2377}\inst{\ref{aff83}}
\and D.~Maino\inst{\ref{aff84},\ref{aff38},\ref{aff85}}
\and E.~Maiorano\orcid{0000-0003-2593-4355}\inst{\ref{aff18}}
\and O.~Mansutti\orcid{0000-0001-5758-4658}\inst{\ref{aff20}}
\and S.~Marcin\inst{\ref{aff86}}
\and O.~Marggraf\orcid{0000-0001-7242-3852}\inst{\ref{aff87}}
\and M.~Martinelli\orcid{0000-0002-6943-7732}\inst{\ref{aff43},\ref{aff88}}
\and N.~Martinet\orcid{0000-0003-2786-7790}\inst{\ref{aff51}}
\and F.~Marulli\orcid{0000-0002-8850-0303}\inst{\ref{aff89},\ref{aff18},\ref{aff24}}
\and R.~J.~Massey\orcid{0000-0002-6085-3780}\inst{\ref{aff90}}
\and N.~Mauri\orcid{0000-0001-8196-1548}\inst{\ref{aff46},\ref{aff24}}
\and E.~Medinaceli\orcid{0000-0002-4040-7783}\inst{\ref{aff18}}
\and S.~Mei\orcid{0000-0002-2849-559X}\inst{\ref{aff91},\ref{aff92}}
\and M.~Melchior\inst{\ref{aff93}}
\and Y.~Mellier\thanks{Deceased}\inst{\ref{aff94},\ref{aff95}}
\and M.~Meneghetti\orcid{0000-0003-1225-7084}\inst{\ref{aff18},\ref{aff24}}
\and E.~Merlin\orcid{0000-0001-6870-8900}\inst{\ref{aff43}}
\and G.~Meylan\inst{\ref{aff96}}
\and P.~Monaco\orcid{0000-0003-2083-7564}\inst{\ref{aff97},\ref{aff20},\ref{aff21},\ref{aff19},\ref{aff98}}
\and A.~Mora\orcid{0000-0002-1922-8529}\inst{\ref{aff99}}
\and M.~Moresco\orcid{0000-0002-7616-7136}\inst{\ref{aff89},\ref{aff18}}
\and C.~Moretti\orcid{0000-0003-3314-8936}\inst{\ref{aff20},\ref{aff19},\ref{aff21},\ref{aff22}}
\and L.~Moscardini\orcid{0000-0002-3473-6716}\inst{\ref{aff89},\ref{aff18},\ref{aff24}}
\and R.~Nakajima\orcid{0009-0009-1213-7040}\inst{\ref{aff87}}
\and C.~Neissner\orcid{0000-0001-8524-4968}\inst{\ref{aff100},\ref{aff40}}
\and S.-M.~Niemi\orcid{0009-0005-0247-0086}\inst{\ref{aff36}}
\and C.~Padilla\orcid{0000-0001-7951-0166}\inst{\ref{aff100}}
\and K.~Paech\orcid{0000-0003-0625-2367}\inst{\ref{aff62}}
\and S.~Paltani\orcid{0000-0002-8108-9179}\inst{\ref{aff58}}
\and F.~Pasian\orcid{0000-0002-4869-3227}\inst{\ref{aff20}}
\and W.~J.~Percival\orcid{0000-0002-0644-5727}\inst{\ref{aff101},\ref{aff102},\ref{aff103}}
\and V.~Pettorino\orcid{0000-0002-4203-9320}\inst{\ref{aff36}}
\and A.~Pezzotta\orcid{0000-0003-0726-2268}\inst{\ref{aff17}}
\and S.~Pires\orcid{0000-0002-0249-2104}\inst{\ref{aff104}}
\and G.~Polenta\orcid{0000-0003-4067-9196}\inst{\ref{aff105}}
\and M.~Poncet\inst{\ref{aff106}}
\and L.~A.~Popa\inst{\ref{aff107}}
\and C.~Porciani\orcid{0000-0002-7797-2508}\inst{\ref{aff87}}
\and L.~Pozzetti\orcid{0000-0001-7085-0412}\inst{\ref{aff18}}
\and F.~Raison\orcid{0000-0002-7819-6918}\inst{\ref{aff62}}
\and A.~Renzi\orcid{0000-0001-9856-1970}\inst{\ref{aff108},\ref{aff60}}
\and J.~Rhodes\orcid{0000-0002-4485-8549}\inst{\ref{aff70}}
\and G.~Riccio\inst{\ref{aff29}}
\and I.~Risso\orcid{0000-0003-2525-7761}\inst{\ref{aff17},\ref{aff27}}
\and E.~Romelli\orcid{0000-0003-3069-9222}\inst{\ref{aff20}}
\and M.~Roncarelli\orcid{0000-0001-9587-7822}\inst{\ref{aff18}}
\and R.~Saglia\orcid{0000-0003-0378-7032}\inst{\ref{aff63},\ref{aff62}}
\and Z.~Sakr\orcid{0000-0002-4823-3757}\inst{\ref{aff109},\ref{aff110},\ref{aff111}}
\and D.~Sapone\orcid{0000-0001-7089-4503}\inst{\ref{aff112}}
\and B.~Sartoris\orcid{0000-0003-1337-5269}\inst{\ref{aff63},\ref{aff20}}
\and P.~Schneider\orcid{0000-0001-8561-2679}\inst{\ref{aff87}}
\and T.~Schrabback\orcid{0000-0002-6987-7834}\inst{\ref{aff113}}
\and A.~Secroun\orcid{0000-0003-0505-3710}\inst{\ref{aff61}}
\and G.~Seidel\orcid{0000-0003-2907-353X}\inst{\ref{aff74}}
\and S.~Serrano\orcid{0000-0002-0211-2861}\inst{\ref{aff114},\ref{aff115},\ref{aff10}}
\and P.~Simon\inst{\ref{aff87}}
\and C.~Sirignano\orcid{0000-0002-0995-7146}\inst{\ref{aff108},\ref{aff60}}
\and G.~Sirri\orcid{0000-0003-2626-2853}\inst{\ref{aff24}}
\and J.~Skottfelt\orcid{0000-0003-1310-8283}\inst{\ref{aff116}}
\and L.~Stanco\orcid{0000-0002-9706-5104}\inst{\ref{aff60}}
\and J.~Steinwagner\orcid{0000-0001-7443-1047}\inst{\ref{aff62}}
\and P.~Tallada-Cresp\'{i}\orcid{0000-0002-1336-8328}\inst{\ref{aff39},\ref{aff40}}
\and A.~N.~Taylor\inst{\ref{aff47}}
\and H.~I.~Teplitz\orcid{0000-0002-7064-5424}\inst{\ref{aff117}}
\and I.~Tereno\orcid{0000-0002-4537-6218}\inst{\ref{aff56},\ref{aff9}}
\and N.~Tessore\orcid{0000-0002-9696-7931}\inst{\ref{aff76}}
\and S.~Toft\orcid{0000-0003-3631-7176}\inst{\ref{aff118},\ref{aff119}}
\and R.~Toledo-Moreo\orcid{0000-0002-2997-4859}\inst{\ref{aff120}}
\and F.~Torradeflot\orcid{0000-0003-1160-1517}\inst{\ref{aff40},\ref{aff39}}
\and I.~Tutusaus\orcid{0000-0002-3199-0399}\inst{\ref{aff10},\ref{aff114},\ref{aff110}}
\and L.~Valenziano\orcid{0000-0002-1170-0104}\inst{\ref{aff18},\ref{aff65}}
\and J.~Valiviita\orcid{0000-0001-6225-3693}\inst{\ref{aff79},\ref{aff80}}
\and T.~Vassallo\orcid{0000-0001-6512-6358}\inst{\ref{aff20}}
\and G.~Verdoes~Kleijn\orcid{0000-0001-5803-2580}\inst{\ref{aff121}}
\and A.~Veropalumbo\orcid{0000-0003-2387-1194}\inst{\ref{aff17},\ref{aff27},\ref{aff26}}
\and Y.~Wang\orcid{0000-0002-4749-2984}\inst{\ref{aff117}}
\and J.~Weller\orcid{0000-0002-8282-2010}\inst{\ref{aff63},\ref{aff62}}
\and G.~Zamorani\orcid{0000-0002-2318-301X}\inst{\ref{aff18}}
\and E.~Zucca\orcid{0000-0002-5845-8132}\inst{\ref{aff18}}
\and V.~Allevato\orcid{0000-0001-7232-5152}\inst{\ref{aff29}}
\and M.~Ballardini\orcid{0000-0003-4481-3559}\inst{\ref{aff122},\ref{aff123},\ref{aff18}}
\and E.~Bozzo\orcid{0000-0002-8201-1525}\inst{\ref{aff58}}
\and C.~Burigana\orcid{0000-0002-3005-5796}\inst{\ref{aff124},\ref{aff65}}
\and R.~Cabanac\orcid{0000-0001-6679-2600}\inst{\ref{aff110}}
\and M.~Calabrese\orcid{0000-0002-2637-2422}\inst{\ref{aff125},\ref{aff38}}
\and A.~Cappi\inst{\ref{aff18},\ref{aff126}}
\and J.~A.~Escartin~Vigo\inst{\ref{aff62}}
\and J.~Mart\'{i}n-Fleitas\orcid{0000-0002-8594-569X}\inst{\ref{aff127}}
\and S.~Matthew\orcid{0000-0001-8448-1697}\inst{\ref{aff47}}
\and R.~B.~Metcalf\orcid{0000-0003-3167-2574}\inst{\ref{aff89},\ref{aff18}}
\and A.~A.~Nucita\inst{\ref{aff128},\ref{aff129},\ref{aff130}}
\and M.~P\"ontinen\orcid{0000-0001-5442-2530}\inst{\ref{aff79}}
\and V.~Scottez\orcid{0009-0008-3864-940X}\inst{\ref{aff94},\ref{aff131}}
\and M.~Sereno\orcid{0000-0003-0302-0325}\inst{\ref{aff18},\ref{aff24}}
\and M.~Tenti\orcid{0000-0002-4254-5901}\inst{\ref{aff24}}
\and M.~Viel\orcid{0000-0002-2642-5707}\inst{\ref{aff19},\ref{aff20},\ref{aff22},\ref{aff21},\ref{aff98}}
\and M.~Wiesmann\orcid{0009-0000-8199-5860}\inst{\ref{aff69}}
\and Y.~Akrami\orcid{0000-0002-2407-7956}\inst{\ref{aff132},\ref{aff133}}
\and I.~T.~Andika\orcid{0000-0001-6102-9526}\inst{\ref{aff134},\ref{aff135}}
\and S.~Anselmi\orcid{0000-0002-3579-9583}\inst{\ref{aff60},\ref{aff108},\ref{aff136}}
\and M.~Archidiacono\orcid{0000-0003-4952-9012}\inst{\ref{aff84},\ref{aff85}}
\and F.~Atrio-Barandela\orcid{0000-0002-2130-2513}\inst{\ref{aff137}}
\and P.~Bergamini\orcid{0000-0003-1383-9414}\inst{\ref{aff18}}
\and D.~Bertacca\orcid{0000-0002-2490-7139}\inst{\ref{aff108},\ref{aff25},\ref{aff60}}
\and M.~Bethermin\orcid{0000-0002-3915-2015}\inst{\ref{aff138}}
\and A.~Blanchard\orcid{0000-0001-8555-9003}\inst{\ref{aff110}}
\and L.~Blot\orcid{0000-0002-9622-7167}\inst{\ref{aff139},\ref{aff81}}
\and M.~Bonici\orcid{0000-0002-8430-126X}\inst{\ref{aff101},\ref{aff38}}
\and M.~L.~Brown\orcid{0000-0002-0370-8077}\inst{\ref{aff48}}
\and S.~Bruton\orcid{0000-0002-6503-5218}\inst{\ref{aff140}}
\and A.~Calabro\orcid{0000-0003-2536-1614}\inst{\ref{aff43}}
\and B.~Camacho~Quevedo\orcid{0000-0002-8789-4232}\inst{\ref{aff19},\ref{aff22},\ref{aff20}}
\and F.~Caro\inst{\ref{aff43}}
\and C.~S.~Carvalho\inst{\ref{aff9}}
\and T.~Castro\orcid{0000-0002-6292-3228}\inst{\ref{aff20},\ref{aff21},\ref{aff19},\ref{aff98}}
\and F.~Cogato\orcid{0000-0003-4632-6113}\inst{\ref{aff89},\ref{aff18}}
\and S.~Conseil\orcid{0000-0002-3657-4191}\inst{\ref{aff50}}
\and T.~Contini\orcid{0000-0003-0275-938X}\inst{\ref{aff110}}
\and A.~R.~Cooray\orcid{0000-0002-3892-0190}\inst{\ref{aff141}}
\and O.~Cucciati\orcid{0000-0002-9336-7551}\inst{\ref{aff18}}
\and G.~Desprez\orcid{0000-0001-8325-1742}\inst{\ref{aff121}}
\and A.~D\'iaz-S\'anchez\orcid{0000-0003-0748-4768}\inst{\ref{aff142}}
\and S.~Di~Domizio\orcid{0000-0003-2863-5895}\inst{\ref{aff26},\ref{aff27}}
\and J.~M.~Diego\orcid{0000-0001-9065-3926}\inst{\ref{aff143}}
\and P.~Dimauro\orcid{0000-0001-7399-2854}\inst{\ref{aff144},\ref{aff43}}
\and P.-A.~Duc\orcid{0000-0003-3343-6284}\inst{\ref{aff138}}
\and M.~Y.~Elkhashab\orcid{0000-0001-9306-2603}\inst{\ref{aff20},\ref{aff21},\ref{aff97},\ref{aff19}}
\and A.~Enia\orcid{0000-0002-0200-2857}\inst{\ref{aff18},\ref{aff23}}
\and Y.~Fang\orcid{0000-0002-0334-6950}\inst{\ref{aff63}}
\and A.~Finoguenov\orcid{0000-0002-4606-5403}\inst{\ref{aff79}}
\and F.~Fontanot\orcid{0000-0003-4744-0188}\inst{\ref{aff20},\ref{aff19}}
\and A.~Franco\orcid{0000-0002-4761-366X}\inst{\ref{aff129},\ref{aff128},\ref{aff130}}
\and K.~Ganga\orcid{0000-0001-8159-8208}\inst{\ref{aff91}}
\and J.~Garc\'ia-Bellido\orcid{0000-0002-9370-8360}\inst{\ref{aff132}}
\and T.~Gasparetto\orcid{0000-0002-7913-4866}\inst{\ref{aff43}}
\and V.~Gautard\inst{\ref{aff145}}
\and R.~Gavazzi\orcid{0000-0002-5540-6935}\inst{\ref{aff51},\ref{aff95}}
\and E.~Gaztanaga\orcid{0000-0001-9632-0815}\inst{\ref{aff10},\ref{aff114},\ref{aff146}}
\and F.~Giacomini\orcid{0000-0002-3129-2814}\inst{\ref{aff24}}
\and F.~Gianotti\orcid{0000-0003-4666-119X}\inst{\ref{aff18}}
\and G.~Gozaliasl\orcid{0000-0002-0236-919X}\inst{\ref{aff147},\ref{aff79}}
\and M.~Guidi\orcid{0000-0001-9408-1101}\inst{\ref{aff23},\ref{aff18}}
\and C.~M.~Gutierrez\orcid{0000-0001-7854-783X}\inst{\ref{aff1},\ref{aff2}}
\and A.~Hall\orcid{0000-0002-3139-8651}\inst{\ref{aff47}}
\and C.~Hern\'andez-Monteagudo\orcid{0000-0001-5471-9166}\inst{\ref{aff2},\ref{aff1}}
\and H.~Hildebrandt\orcid{0000-0002-9814-3338}\inst{\ref{aff148}}
\and J.~Hjorth\orcid{0000-0002-4571-2306}\inst{\ref{aff149}}
\and J.~J.~E.~Kajava\orcid{0000-0002-3010-8333}\inst{\ref{aff150},\ref{aff151}}
\and Y.~Kang\orcid{0009-0000-8588-7250}\inst{\ref{aff58}}
\and V.~Kansal\orcid{0000-0002-4008-6078}\inst{\ref{aff152},\ref{aff153}}
\and D.~Karagiannis\orcid{0000-0002-4927-0816}\inst{\ref{aff122},\ref{aff154}}
\and J.~Kim\orcid{0000-0003-2776-2761}\inst{\ref{aff155}}
\and C.~C.~Kirkpatrick\inst{\ref{aff77}}
\and S.~Kruk\orcid{0000-0001-8010-8879}\inst{\ref{aff15}}
\and L.~Legrand\orcid{0000-0003-0610-5252}\inst{\ref{aff156},\ref{aff157}}
\and M.~Lembo\orcid{0000-0002-5271-5070}\inst{\ref{aff95},\ref{aff122},\ref{aff123}}
\and F.~Lepori\orcid{0009-0000-5061-7138}\inst{\ref{aff158}}
\and G.~Leroy\orcid{0009-0004-2523-4425}\inst{\ref{aff159},\ref{aff90}}
\and G.~F.~Lesci\orcid{0000-0002-4607-2830}\inst{\ref{aff89},\ref{aff18}}
\and J.~Lesgourgues\orcid{0000-0001-7627-353X}\inst{\ref{aff41}}
\and L.~Leuzzi\orcid{0009-0006-4479-7017}\inst{\ref{aff18}}
\and T.~I.~Liaudat\orcid{0000-0002-9104-314X}\inst{\ref{aff160}}
\and A.~Loureiro\orcid{0000-0002-4371-0876}\inst{\ref{aff161},\ref{aff162}}
\and J.~Macias-Perez\orcid{0000-0002-5385-2763}\inst{\ref{aff163}}
\and E.~A.~Magnier\orcid{0000-0002-7965-2815}\inst{\ref{aff45}}
\and F.~Mannucci\orcid{0000-0002-4803-2381}\inst{\ref{aff164}}
\and R.~Maoli\orcid{0000-0002-6065-3025}\inst{\ref{aff165},\ref{aff43}}
\and C.~J.~A.~P.~Martins\orcid{0000-0002-4886-9261}\inst{\ref{aff166},\ref{aff30}}
\and L.~Maurin\orcid{0000-0002-8406-0857}\inst{\ref{aff59}}
\and G.~Morgante\inst{\ref{aff18}}
\and K.~Naidoo\orcid{0000-0002-9182-1802}\inst{\ref{aff146},\ref{aff76}}
\and P.~Natoli\orcid{0000-0003-0126-9100}\inst{\ref{aff122},\ref{aff123}}
\and A.~Navarro-Alsina\orcid{0000-0002-3173-2592}\inst{\ref{aff87}}
\and S.~Nesseris\orcid{0000-0002-0567-0324}\inst{\ref{aff132}}
\and D.~Paoletti\orcid{0000-0003-4761-6147}\inst{\ref{aff18},\ref{aff65}}
\and F.~Passalacqua\orcid{0000-0002-8606-4093}\inst{\ref{aff108},\ref{aff60}}
\and K.~Paterson\orcid{0000-0001-8340-3486}\inst{\ref{aff74}}
\and A.~Pisani\orcid{0000-0002-6146-4437}\inst{\ref{aff61}}
\and D.~Potter\orcid{0000-0002-0757-5195}\inst{\ref{aff158}}
\and S.~Quai\orcid{0000-0002-0449-8163}\inst{\ref{aff89},\ref{aff18}}
\and M.~Radovich\orcid{0000-0002-3585-866X}\inst{\ref{aff25}}
\and G.~Rodighiero\orcid{0000-0002-9415-2296}\inst{\ref{aff108},\ref{aff25}}
\and S.~Sacquegna\orcid{0000-0002-8433-6630}\inst{\ref{aff167},\ref{aff128},\ref{aff129}}
\and M.~Sahl\'en\orcid{0000-0003-0973-4804}\inst{\ref{aff168}}
\and D.~B.~Sanders\orcid{0000-0002-1233-9998}\inst{\ref{aff45}}
\and E.~Sarpa\orcid{0000-0002-1256-655X}\inst{\ref{aff22},\ref{aff98},\ref{aff21}}
\and A.~Schneider\orcid{0000-0001-7055-8104}\inst{\ref{aff158}}
\and M.~Schultheis\inst{\ref{aff126}}
\and D.~Sciotti\orcid{0009-0008-4519-2620}\inst{\ref{aff43},\ref{aff88}}
\and E.~Sellentin\inst{\ref{aff169},\ref{aff37}}
\and F.~Shankar\orcid{0000-0001-8973-5051}\inst{\ref{aff170}}
\and L.~C.~Smith\orcid{0000-0002-3259-2771}\inst{\ref{aff171}}
\and J.~G.~Sorce\orcid{0000-0002-2307-2432}\inst{\ref{aff172},\ref{aff59}}
\and K.~Tanidis\orcid{0000-0001-9843-5130}\inst{\ref{aff155}}
\and G.~Testera\inst{\ref{aff27}}
\and R.~Teyssier\orcid{0000-0001-7689-0933}\inst{\ref{aff173}}
\and S.~Tosi\orcid{0000-0002-7275-9193}\inst{\ref{aff26},\ref{aff27},\ref{aff17}}
\and A.~Troja\orcid{0000-0003-0239-4595}\inst{\ref{aff108},\ref{aff60}}
\and M.~Tucci\inst{\ref{aff58}}
\and C.~Valieri\inst{\ref{aff24}}
\and A.~Venhola\orcid{0000-0001-6071-4564}\inst{\ref{aff174}}
\and D.~Vergani\orcid{0000-0003-0898-2216}\inst{\ref{aff18}}
\and G.~Verza\orcid{0000-0002-1886-8348}$^{\star\star}$\inst{\ref{aff175}}
\and P.~Vielzeuf\orcid{0000-0003-2035-9339}\inst{\ref{aff61}}
\and N.~A.~Walton\orcid{0000-0003-3983-8778}\inst{\ref{aff171}}
\and D.~Scott\orcid{0000-0002-6878-9840}\inst{\ref{aff176}}}
										   
%%%% please do not edit the affiliation list -- contact ECEB Bureau for changes
\institute{Instituto de Astrof\'{\i}sica de Canarias, E-38205 La Laguna, Tenerife, Spain\label{aff1}
\and
Universidad de La Laguna, Dpto. Astrof\'\i sica, E-38206 La Laguna, Tenerife, Spain\label{aff2}
\and
NASA Ames Research Center, Moffett Field, CA 94035, USA\label{aff11}
\and
Departamento de Fisica, Universidad de Cordoba, Campus de Rabanales, Edificio Albert Einstein, E-14071 Cordoba, Spain\label{aff3}
\and
Universit\'e PSL, Observatoire de Paris, Sorbonne Universit\'e, CNRS, LERMA, 75014, Paris, France\label{aff5}
\and
Universit\'e Paris-Cit\'e, 5 Rue Thomas Mann, 75013, Paris, France\label{aff6}
\and
Departamento de F\'{i}sica Te\'{o}rica, At\'{o}mica y \'{O}ptica, Universidad de Valladolid, 47011 Valladolid, Spain\label{aff7}
\and
Laboratory for Disruptive Interdisciplinary Science (LaDIS), Universidad de Valladolid, 47011 Valladolid, Spain\label{aff8}
\and
Instituto de Astrof\'isica e Ci\^encias do Espa\c{c}o, Faculdade de Ci\^encias, Universidade de Lisboa, Tapada da Ajuda, 1349-018 Lisboa, Portugal\label{aff9}
\and
Institute of Space Sciences (ICE, CSIC), Campus UAB, Carrer de Can Magrans, s/n, 08193 Barcelona, Spain\label{aff10}
\and
Bay Area Environmental Research Institute, Moffett Field, California 94035, USA\label{aff12}
\and
Leibniz-Institut f\"{u}r Astrophysik (AIP), An der Sternwarte 16, 14482 Potsdam, Germany\label{aff13}
\and
Bernoulli Institute for Mathematics, Computer Science and Artificial Intelligence, University of Groningen, PO Box 407, 9700 AK Groningen, The Netherlands\label{aff14}
\and
ESAC/ESA, Camino Bajo del Castillo, s/n., Urb. Villafranca del Castillo, 28692 Villanueva de la Ca\~nada, Madrid, Spain\label{aff15}
\and
School of Mathematics and Physics, University of Surrey, Guildford, Surrey, GU2 7XH, UK\label{aff16}
\and
INAF-Osservatorio Astronomico di Brera, Via Brera 28, 20122 Milano, Italy\label{aff17}
\and
INAF-Osservatorio di Astrofisica e Scienza dello Spazio di Bologna, Via Piero Gobetti 93/3, 40129 Bologna, Italy\label{aff18}
\and
IFPU, Institute for Fundamental Physics of the Universe, via Beirut 2, 34151 Trieste, Italy\label{aff19}
\and
INAF-Osservatorio Astronomico di Trieste, Via G. B. Tiepolo 11, 34143 Trieste, Italy\label{aff20}
\and
INFN, Sezione di Trieste, Via Valerio 2, 34127 Trieste TS, Italy\label{aff21}
\and
SISSA, International School for Advanced Studies, Via Bonomea 265, 34136 Trieste TS, Italy\label{aff22}
\and
Dipartimento di Fisica e Astronomia, Universit\`a di Bologna, Via Gobetti 93/2, 40129 Bologna, Italy\label{aff23}
\and
INFN-Sezione di Bologna, Viale Berti Pichat 6/2, 40127 Bologna, Italy\label{aff24}
\and
INAF-Osservatorio Astronomico di Padova, Via dell'Osservatorio 5, 35122 Padova, Italy\label{aff25}
\and
Dipartimento di Fisica, Universit\`a di Genova, Via Dodecaneso 33, 16146, Genova, Italy\label{aff26}
\and
INFN-Sezione di Genova, Via Dodecaneso 33, 16146, Genova, Italy\label{aff27}
\and
Department of Physics "E. Pancini", University Federico II, Via Cinthia 6, 80126, Napoli, Italy\label{aff28}
\and
INAF-Osservatorio Astronomico di Capodimonte, Via Moiariello 16, 80131 Napoli, Italy\label{aff29}
\and
Instituto de Astrof\'isica e Ci\^encias do Espa\c{c}o, Universidade do Porto, CAUP, Rua das Estrelas, PT4150-762 Porto, Portugal\label{aff30}
\and
Faculdade de Ci\^encias da Universidade do Porto, Rua do Campo de Alegre, 4150-007 Porto, Portugal\label{aff31}
\and
European Southern Observatory, Karl-Schwarzschild-Str.~2, 85748 Garching, Germany\label{aff32}
\and
Dipartimento di Fisica, Universit\`a degli Studi di Torino, Via P. Giuria 1, 10125 Torino, Italy\label{aff33}
\and
INFN-Sezione di Torino, Via P. Giuria 1, 10125 Torino, Italy\label{aff34}
\and
INAF-Osservatorio Astrofisico di Torino, Via Osservatorio 20, 10025 Pino Torinese (TO), Italy\label{aff35}
\and
European Space Agency/ESTEC, Keplerlaan 1, 2201 AZ Noordwijk, The Netherlands\label{aff36}
\and
Leiden Observatory, Leiden University, Einsteinweg 55, 2333 CC Leiden, The Netherlands\label{aff37}
\and
INAF-IASF Milano, Via Alfonso Corti 12, 20133 Milano, Italy\label{aff38}
\and
Centro de Investigaciones Energ\'eticas, Medioambientales y Tecnol\'ogicas (CIEMAT), Avenida Complutense 40, 28040 Madrid, Spain\label{aff39}
\and
Port d'Informaci\'{o} Cient\'{i}fica, Campus UAB, C. Albareda s/n, 08193 Bellaterra (Barcelona), Spain\label{aff40}
\and
Institute for Theoretical Particle Physics and Cosmology (TTK), RWTH Aachen University, 52056 Aachen, Germany\label{aff41}
\and
Deutsches Zentrum f\"ur Luft- und Raumfahrt e. V. (DLR), Linder H\"ohe, 51147 K\"oln, Germany\label{aff42}
\and
INAF-Osservatorio Astronomico di Roma, Via Frascati 33, 00078 Monteporzio Catone, Italy\label{aff43}
\and
INFN section of Naples, Via Cinthia 6, 80126, Napoli, Italy\label{aff44}
\and
Institute for Astronomy, University of Hawaii, 2680 Woodlawn Drive, Honolulu, HI 96822, USA\label{aff45}
\and
Dipartimento di Fisica e Astronomia "Augusto Righi" - Alma Mater Studiorum Universit\`a di Bologna, Viale Berti Pichat 6/2, 40127 Bologna, Italy\label{aff46}
\and
Institute for Astronomy, University of Edinburgh, Royal Observatory, Blackford Hill, Edinburgh EH9 3HJ, UK\label{aff47}
\and
Jodrell Bank Centre for Astrophysics, Department of Physics and Astronomy, University of Manchester, Oxford Road, Manchester M13 9PL, UK\label{aff48}
\and
European Space Agency/ESRIN, Largo Galileo Galilei 1, 00044 Frascati, Roma, Italy\label{aff49}
\and
Universit\'e Claude Bernard Lyon 1, CNRS/IN2P3, IP2I Lyon, UMR 5822, Villeurbanne, F-69100, France\label{aff50}
\and
Aix-Marseille Universit\'e, CNRS, CNES, LAM, Marseille, France\label{aff51}
\and
Institut de Ci\`{e}ncies del Cosmos (ICCUB), Universitat de Barcelona (IEEC-UB), Mart\'{i} i Franqu\`{e}s 1, 08028 Barcelona, Spain\label{aff52}
\and
Instituci\'o Catalana de Recerca i Estudis Avan\c{c}ats (ICREA), Passeig de Llu\'{\i}s Companys 23, 08010 Barcelona, Spain\label{aff53}
\and
UCB Lyon 1, CNRS/IN2P3, IUF, IP2I Lyon, 4 rue Enrico Fermi, 69622 Villeurbanne, France\label{aff54}
\and
Mullard Space Science Laboratory, University College London, Holmbury St Mary, Dorking, Surrey RH5 6NT, UK\label{aff55}
\and
Departamento de F\'isica, Faculdade de Ci\^encias, Universidade de Lisboa, Edif\'icio C8, Campo Grande, PT1749-016 Lisboa, Portugal\label{aff56}
\and
Instituto de Astrof\'isica e Ci\^encias do Espa\c{c}o, Faculdade de Ci\^encias, Universidade de Lisboa, Campo Grande, 1749-016 Lisboa, Portugal\label{aff57}
\and
Department of Astronomy, University of Geneva, ch. d'Ecogia 16, 1290 Versoix, Switzerland\label{aff58}
\and
Universit\'e Paris-Saclay, CNRS, Institut d'astrophysique spatiale, 91405, Orsay, France\label{aff59}
\and
INFN-Padova, Via Marzolo 8, 35131 Padova, Italy\label{aff60}
\and
Aix-Marseille Universit\'e, CNRS/IN2P3, CPPM, Marseille, France\label{aff61}
\and
Max Planck Institute for Extraterrestrial Physics, Giessenbachstr. 1, 85748 Garching, Germany\label{aff62}
\and
Universit\"ats-Sternwarte M\"unchen, Fakult\"at f\"ur Physik, Ludwig-Maximilians-Universit\"at M\"unchen, Scheinerstr.~1, 81679 M\"unchen, Germany\label{aff63}
\and
INAF-Istituto di Astrofisica e Planetologia Spaziali, via del Fosso del Cavaliere, 100, 00100 Roma, Italy\label{aff64}
\and
INFN-Bologna, Via Irnerio 46, 40126 Bologna, Italy\label{aff65}
\and
School of Physics, HH Wills Physics Laboratory, University of Bristol, Tyndall Avenue, Bristol, BS8 1TL, UK\label{aff66}
\and
University Observatory, LMU Faculty of Physics, Scheinerstr.~1, 81679 Munich, Germany\label{aff67}
\and
National Research Council, Herzberg Astronomy and Astrophysics Research Centre, 5071 W. Saanich Rd. Victoria, BC, V9E 2E7, Canada\label{aff68}
\and
Institute of Theoretical Astrophysics, University of Oslo, P.O. Box 1029 Blindern, 0315 Oslo, Norway\label{aff69}
\and
Jet Propulsion Laboratory, California Institute of Technology, 4800 Oak Grove Drive, Pasadena, CA, 91109, USA\label{aff70}
\and
Felix Hormuth Engineering, Goethestr. 17, 69181 Leimen, Germany\label{aff71}
\and
Technical University of Denmark, Elektrovej 327, 2800 Kgs. Lyngby, Denmark\label{aff72}
\and
Cosmic Dawn Center (DAWN), Denmark\label{aff73}
\and
Max-Planck-Institut f\"ur Astronomie, K\"onigstuhl 17, 69117 Heidelberg, Germany\label{aff74}
\and
NASA Goddard Space Flight Center, Greenbelt, MD 20771, USA\label{aff75}
\and
Department of Physics and Astronomy, University College London, Gower Street, London WC1E 6BT, UK\label{aff76}
\and
Department of Physics and Helsinki Institute of Physics, Gustaf H\"allstr\"omin katu 2, University of Helsinki, 00014 Helsinki, Finland\label{aff77}
\and
Universit\'e de Gen\`eve, D\'epartement de Physique Th\'eorique and Centre for Astroparticle Physics, 24 quai Ernest-Ansermet, CH-1211 Gen\`eve 4, Switzerland\label{aff78}
\and
Department of Physics, P.O. Box 64, University of Helsinki, 00014 Helsinki, Finland\label{aff79}
\and
Helsinki Institute of Physics, Gustaf H{\"a}llstr{\"o}min katu 2, University of Helsinki, 00014 Helsinki, Finland\label{aff80}
\and
Laboratoire d'etude de l'Univers et des phenomenes eXtremes, Observatoire de Paris, Universit\'e PSL, Sorbonne Universit\'e, CNRS, 92190 Meudon, France\label{aff81}
\and
SKAO, Jodrell Bank, Lower Withington, Macclesfield SK11 9FT, UK\label{aff82}
\and
Centre de Calcul de l'IN2P3/CNRS, 21 avenue Pierre de Coubertin 69627 Villeurbanne Cedex, France\label{aff83}
\and
Dipartimento di Fisica "Aldo Pontremoli", Universit\`a degli Studi di Milano, Via Celoria 16, 20133 Milano, Italy\label{aff84}
\and
INFN-Sezione di Milano, Via Celoria 16, 20133 Milano, Italy\label{aff85}
\and
University of Applied Sciences and Arts of Northwestern Switzerland, School of Computer Science, 5210 Windisch, Switzerland\label{aff86}
\and
Universit\"at Bonn, Argelander-Institut f\"ur Astronomie, Auf dem H\"ugel 71, 53121 Bonn, Germany\label{aff87}
\and
INFN-Sezione di Roma, Piazzale Aldo Moro, 2 - c/o Dipartimento di Fisica, Edificio G. Marconi, 00185 Roma, Italy\label{aff88}
\and
Dipartimento di Fisica e Astronomia "Augusto Righi" - Alma Mater Studiorum Universit\`a di Bologna, via Piero Gobetti 93/2, 40129 Bologna, Italy\label{aff89}
\and
Department of Physics, Institute for Computational Cosmology, Durham University, South Road, Durham, DH1 3LE, UK\label{aff90}
\and
Universit\'e Paris Cit\'e, CNRS, Astroparticule et Cosmologie, 75013 Paris, France\label{aff91}
\and
CNRS-UCB International Research Laboratory, Centre Pierre Bin\'etruy, IRL2007, CPB-IN2P3, Berkeley, USA\label{aff92}
\and
University of Applied Sciences and Arts of Northwestern Switzerland, School of Engineering, 5210 Windisch, Switzerland\label{aff93}
\and
Institut d'Astrophysique de Paris, 98bis Boulevard Arago, 75014, Paris, France\label{aff94}
\and
Institut d'Astrophysique de Paris, UMR 7095, CNRS, and Sorbonne Universit\'e, 98 bis boulevard Arago, 75014 Paris, France\label{aff95}
\and
Institute of Physics, Laboratory of Astrophysics, Ecole Polytechnique F\'ed\'erale de Lausanne (EPFL), Observatoire de Sauverny, 1290 Versoix, Switzerland\label{aff96}
\and
Dipartimento di Fisica - Sezione di Astronomia, Universit\`a di Trieste, Via Tiepolo 11, 34131 Trieste, Italy\label{aff97}
\and
ICSC - Centro Nazionale di Ricerca in High Performance Computing, Big Data e Quantum Computing, Via Magnanelli 2, Bologna, Italy\label{aff98}
\and
Telespazio UK S.L. for European Space Agency (ESA), Camino bajo del Castillo, s/n, Urbanizacion Villafranca del Castillo, Villanueva de la Ca\~nada, 28692 Madrid, Spain\label{aff99}
\and
Institut de F\'{i}sica d'Altes Energies (IFAE), The Barcelona Institute of Science and Technology, Campus UAB, 08193 Bellaterra (Barcelona), Spain\label{aff100}
\and
Waterloo Centre for Astrophysics, University of Waterloo, Waterloo, Ontario N2L 3G1, Canada\label{aff101}
\and
Department of Physics and Astronomy, University of Waterloo, Waterloo, Ontario N2L 3G1, Canada\label{aff102}
\and
Perimeter Institute for Theoretical Physics, Waterloo, Ontario N2L 2Y5, Canada\label{aff103}
\and
Universit\'e Paris-Saclay, Universit\'e Paris Cit\'e, CEA, CNRS, AIM, 91191, Gif-sur-Yvette, France\label{aff104}
\and
Space Science Data Center, Italian Space Agency, via del Politecnico snc, 00133 Roma, Italy\label{aff105}
\and
Centre National d'Etudes Spatiales -- Centre spatial de Toulouse, 18 avenue Edouard Belin, 31401 Toulouse Cedex 9, France\label{aff106}
\and
Institute of Space Science, Str. Atomistilor, nr. 409 M\u{a}gurele, Ilfov, 077125, Romania\label{aff107}
\and
Dipartimento di Fisica e Astronomia "G. Galilei", Universit\`a di Padova, Via Marzolo 8, 35131 Padova, Italy\label{aff108}
\and
Institut f\"ur Theoretische Physik, University of Heidelberg, Philosophenweg 16, 69120 Heidelberg, Germany\label{aff109}
\and
Institut de Recherche en Astrophysique et Plan\'etologie (IRAP), Universit\'e de Toulouse, CNRS, UPS, CNES, 14 Av. Edouard Belin, 31400 Toulouse, France\label{aff110}
\and
Universit\'e St Joseph; Faculty of Sciences, Beirut, Lebanon\label{aff111}
\and
Departamento de F\'isica, FCFM, Universidad de Chile, Blanco Encalada 2008, Santiago, Chile\label{aff112}
\and
Universit\"at Innsbruck, Institut f\"ur Astro- und Teilchenphysik, Technikerstr. 25/8, 6020 Innsbruck, Austria\label{aff113}
\and
Institut d'Estudis Espacials de Catalunya (IEEC),  Edifici RDIT, Campus UPC, 08860 Castelldefels, Barcelona, Spain\label{aff114}
\and
Satlantis, University Science Park, Sede Bld 48940, Leioa-Bilbao, Spain\label{aff115}
\and
Centre for Electronic Imaging, Open University, Walton Hall, Milton Keynes, MK7~6AA, UK\label{aff116}
\and
Infrared Processing and Analysis Center, California Institute of Technology, Pasadena, CA 91125, USA\label{aff117}
\and
Cosmic Dawn Center (DAWN)\label{aff118}
\and
Niels Bohr Institute, University of Copenhagen, Jagtvej 128, 2200 Copenhagen, Denmark\label{aff119}
\and
Universidad Polit\'ecnica de Cartagena, Departamento de Electr\'onica y Tecnolog\'ia de Computadoras,  Plaza del Hospital 1, 30202 Cartagena, Spain\label{aff120}
\and
Kapteyn Astronomical Institute, University of Groningen, PO Box 800, 9700 AV Groningen, The Netherlands\label{aff121}
\and
Dipartimento di Fisica e Scienze della Terra, Universit\`a degli Studi di Ferrara, Via Giuseppe Saragat 1, 44122 Ferrara, Italy\label{aff122}
\and
Istituto Nazionale di Fisica Nucleare, Sezione di Ferrara, Via Giuseppe Saragat 1, 44122 Ferrara, Italy\label{aff123}
\and
INAF, Istituto di Radioastronomia, Via Piero Gobetti 101, 40129 Bologna, Italy\label{aff124}
\and
Astronomical Observatory of the Autonomous Region of the Aosta Valley (OAVdA), Loc. Lignan 39, I-11020, Nus (Aosta Valley), Italy\label{aff125}
\and
Universit\'e C\^{o}te d'Azur, Observatoire de la C\^{o}te d'Azur, CNRS, Laboratoire Lagrange, Bd de l'Observatoire, CS 34229, 06304 Nice cedex 4, France\label{aff126}
\and
Aurora Technology for European Space Agency (ESA), Camino bajo del Castillo, s/n, Urbanizacion Villafranca del Castillo, Villanueva de la Ca\~nada, 28692 Madrid, Spain\label{aff127}
\and
Department of Mathematics and Physics E. De Giorgi, University of Salento, Via per Arnesano, CP-I93, 73100, Lecce, Italy\label{aff128}
\and
INFN, Sezione di Lecce, Via per Arnesano, CP-193, 73100, Lecce, Italy\label{aff129}
\and
INAF-Sezione di Lecce, c/o Dipartimento Matematica e Fisica, Via per Arnesano, 73100, Lecce, Italy\label{aff130}
\and
ICL, Junia, Universit\'e Catholique de Lille, LITL, 59000 Lille, France\label{aff131}
\and
Instituto de F\'isica Te\'orica UAM-CSIC, Campus de Cantoblanco, 28049 Madrid, Spain\label{aff132}
\and
CERCA/ISO, Department of Physics, Case Western Reserve University, 10900 Euclid Avenue, Cleveland, OH 44106, USA\label{aff133}
\and
Technical University of Munich, TUM School of Natural Sciences, Physics Department, James-Franck-Str.~1, 85748 Garching, Germany\label{aff134}
\and
Max-Planck-Institut f\"ur Astrophysik, Karl-Schwarzschild-Str.~1, 85748 Garching, Germany\label{aff135}
\and
Laboratoire Univers et Th\'eorie, Observatoire de Paris, Universit\'e PSL, Universit\'e Paris Cit\'e, CNRS, 92190 Meudon, France\label{aff136}
\and
Departamento de F{\'\i}sica Fundamental. Universidad de Salamanca. Plaza de la Merced s/n. 37008 Salamanca, Spain\label{aff137}
\and
Universit\'e de Strasbourg, CNRS, Observatoire astronomique de Strasbourg, UMR 7550, 67000 Strasbourg, France\label{aff138}
\and
Center for Data-Driven Discovery, Kavli IPMU (WPI), UTIAS, The University of Tokyo, Kashiwa, Chiba 277-8583, Japan\label{aff139}
\and
California Institute of Technology, 1200 E California Blvd, Pasadena, CA 91125, USA\label{aff140}
\and
Department of Physics \& Astronomy, University of California Irvine, Irvine CA 92697, USA\label{aff141}
\and
Departamento F\'isica Aplicada, Universidad Polit\'ecnica de Cartagena, Campus Muralla del Mar, 30202 Cartagena, Murcia, Spain\label{aff142}
\and
Instituto de F\'isica de Cantabria, Edificio Juan Jord\'a, Avenida de los Castros, 39005 Santander, Spain\label{aff143}
\and
Observatorio Nacional, Rua General Jose Cristino, 77-Bairro Imperial de Sao Cristovao, Rio de Janeiro, 20921-400, Brazil\label{aff144}
\and
CEA Saclay, DFR/IRFU, Service d'Astrophysique, Bat. 709, 91191 Gif-sur-Yvette, France\label{aff145}
\and
Institute of Cosmology and Gravitation, University of Portsmouth, Portsmouth PO1 3FX, UK\label{aff146}
\and
Department of Computer Science, Aalto University, PO Box 15400, Espoo, FI-00 076, Finland\label{aff147}
\and
Ruhr University Bochum, Faculty of Physics and Astronomy, Astronomical Institute (AIRUB), German Centre for Cosmological Lensing (GCCL), 44780 Bochum, Germany\label{aff148}
\and
DARK, Niels Bohr Institute, University of Copenhagen, Jagtvej 155, 2200 Copenhagen, Denmark\label{aff149}
\and
Department of Physics and Astronomy, Vesilinnantie 5, University of Turku, 20014 Turku, Finland\label{aff150}
\and
Serco for European Space Agency (ESA), Camino bajo del Castillo, s/n, Urbanizacion Villafranca del Castillo, Villanueva de la Ca\~nada, 28692 Madrid, Spain\label{aff151}
\and
ARC Centre of Excellence for Dark Matter Particle Physics, Melbourne, Australia\label{aff152}
\and
Centre for Astrophysics \& Supercomputing, Swinburne University of Technology,  Hawthorn, Victoria 3122, Australia\label{aff153}
\and
Department of Physics and Astronomy, University of the Western Cape, Bellville, Cape Town, 7535, South Africa\label{aff154}
\and
Department of Physics, Oxford University, Keble Road, Oxford OX1 3RH, UK\label{aff155}
\and
DAMTP, Centre for Mathematical Sciences, Wilberforce Road, Cambridge CB3 0WA, UK\label{aff156}
\and
Kavli Institute for Cosmology Cambridge, Madingley Road, Cambridge, CB3 0HA, UK\label{aff157}
\and
Department of Astrophysics, University of Zurich, Winterthurerstrasse 190, 8057 Zurich, Switzerland\label{aff158}
\and
Department of Physics, Centre for Extragalactic Astronomy, Durham University, South Road, Durham, DH1 3LE, UK\label{aff159}
\and
IRFU, CEA, Universit\'e Paris-Saclay 91191 Gif-sur-Yvette Cedex, France\label{aff160}
\and
Oskar Klein Centre for Cosmoparticle Physics, Department of Physics, Stockholm University, Stockholm, SE-106 91, Sweden\label{aff161}
\and
Astrophysics Group, Blackett Laboratory, Imperial College London, London SW7 2AZ, UK\label{aff162}
\and
Univ. Grenoble Alpes, CNRS, Grenoble INP, LPSC-IN2P3, 53, Avenue des Martyrs, 38000, Grenoble, France\label{aff163}
\and
INAF-Osservatorio Astrofisico di Arcetri, Largo E. Fermi 5, 50125, Firenze, Italy\label{aff164}
\and
Dipartimento di Fisica, Sapienza Universit\`a di Roma, Piazzale Aldo Moro 2, 00185 Roma, Italy\label{aff165}
\and
Centro de Astrof\'{\i}sica da Universidade do Porto, Rua das Estrelas, 4150-762 Porto, Portugal\label{aff166}
\and
INAF - Osservatorio Astronomico d'Abruzzo, Via Maggini, 64100, Teramo, Italy\label{aff167}
\and
Theoretical astrophysics, Department of Physics and Astronomy, Uppsala University, Box 516, 751 37 Uppsala, Sweden\label{aff168}
\and
Mathematical Institute, University of Leiden, Einsteinweg 55, 2333 CA Leiden, The Netherlands\label{aff169}
\and
School of Physics \& Astronomy, University of Southampton, Highfield Campus, Southampton SO17 1BJ, UK\label{aff170}
\and
Institute of Astronomy, University of Cambridge, Madingley Road, Cambridge CB3 0HA, UK\label{aff171}
\and
Univ. Lille, CNRS, Centrale Lille, UMR 9189 CRIStAL, 59000 Lille, France\label{aff172}
\and
Department of Astrophysical Sciences, Peyton Hall, Princeton University, Princeton, NJ 08544, USA\label{aff173}
\and
Space physics and astronomy research unit, University of Oulu, Pentti Kaiteran katu 1, FI-90014 Oulu, Finland\label{aff174}
\and
Center for Computational Astrophysics, Flatiron Institute, 162 5th Avenue, 10010, New York, NY, USA\label{aff175}
\and
Department of Physics and Astronomy, University of British Columbia, Vancouver, BC V6T 1Z1, Canada\label{aff176}}    

\abstract{
Changes in the slope of the radial surface brightness profiles of disc galaxies, or disc breaks, encode valuable information about the processes driving disc formation and growth. Until now, due to the resolution and depth required, statistical studies of break types have been limited mostly to the local Universe.
We aim to study the structural properties and redshift evolution of disc breaks in a large and statistically representative sample of galaxies out to redshift $z=1$  using the superb depth and resolution of \Euclid. 
We develop an automatic pipeline to extract and analyse surface brightness profiles of 8748 disc galaxies selected from the Euclid Quick Release 1 (Q1) dataset. The pipeline includes advanced masking, surface brightness profile extraction, integrated photometry and structural parameter measurements, and piecewise modelling to classify galaxies into Type~I (pure exponential), Type~II (down-bending), Type~III (up-bending), and composite break (e.g., Type~II+III, Type~II+II, and all the possible combinations) profiles. We quantify structural parameters such as scale lengths, break radii, and surface brightness levels at the break radius, and analyse their distributions and redshift dependence.
Our automatic method produces reliable masks and deep surface brightness profiles, and classifies galaxy profiles with an estimated accuracy of 70\%. Using a subsample of 4385 galaxies with reliable profile type classifications, we find that the most common profiles are Type~II, Type~II+III, Type~I, and Type~III.
The break radius correlates tightly with the outer disc scale length, following $r_\mathrm{b}/h_\mathrm{out} \sim 2.9$ (where $r_\mathrm{b}$ is the break radius and $h_\mathrm{out}$ is the outer disc scale length), irrespective of break type. Type~II breaks occur at a brighter surface brightness levels than in Type~III profiles, and composite profiles exhibit an enhanced break strength (defined as the ratio of the inner and outer scale length). We find substantial evolution in the fraction of disc breaks with time. Type~II profiles dominate at low redshift (50\%), while Type~III and Type~II+III profiles increase in frequency with lookback time, reaching $\sim30\%$ each at $z\sim1$. 
Our study provides the first statistically robust measurement of the evolution of disc break types over $\sim5$\,Gyr. The observed trends support a scenario in which secular processes replace environmental influences as galaxies mature. \textit{Euclid}’s depth, resolution, and coverage has opened a new window for understanding galaxy structure formation and evolution.}
%    from the list in
%     https://www.aanda.org/for-authors/latex-issues/information-files#pop}
%
% Add short versions of title and author list for page headings
%
    \titlerunning{Evolution of disc breaks}
    \authorrunning{Euclid Collaboration: P.~M.~Sánchez-Alarcón et al.}
   
   \maketitle
%
%-------------------------------------------------------------------
%
%
%   Start the main text of your paper here
%
   
\section{\label{sc:Intro}Introduction}
%
%Disc galaxies are one of the most common types of galaxies in the Universe \citep{Conselice03, CEERS}. 
Most of the baryonic component and angular momentum of galaxies are found in discs, and this is also where most of the galaxy evolutionary activity, such as the formation of stars, spiral arms, bars, rings, and the various forms of secular evolution \citep{Kormendy04,vanderKruit11, Kormendy13}, also takes place in discs. Hence, studying the radial light distribution of discs can provide crucial insights into their structures, their formation, and their evolution.

It is now well established that the radial surface brightness profiles of disc galaxies often show changes in slope. Therefore, most of the profiles can be described as broken exponentials \citep[e.g.][]{Erwin05, Laine14,Watkins19}. Since the seminal work by \cite{Pohlen06} and \cite{Erwin08}, disc galaxy surface brightness profiles are separated into three distinct categories: Type~I profiles, corresponding to single exponential discs, follow an exponential law spanning several scale~lengths without any break signature; Type~II profiles, corresponding to down-bending discs, are best described as a broken exponential with a steeper decline in the outer slope; and Type~III profiles, corresponding to up-bending discs, are best described with a broken exponential with a shallower decline in the outer slope. 

Only one third of galaxies show Type~I profiles, while Type~II profiles are the most frequent among discs in the local Universe \citep{Pohlen06, Erwin08, Laine14, Pranger17}. However, the fraction of types of breaks varies with respect to the host morphological type \citep{Gutierrez11, Munoz-Mateos13,Laine14}, stellar mass \citep{Laine14}, and environment \citep{Watkins19, Sanchez-Alarcon23}. Undoubtedly, these variations are associated with the origin of the galaxy and its evolution through time. 

Type~II disc breaks are more frequent in late morphological types and high-mass galaxies \citep[$M_*>10^{10}\, M_\odot$,][]{Laine14}. Their origin is commonly associated either with star-formation thresholds \citep{Kennicutt1989, Sanchez-Blazquez09, 2022MNRAS.514.5340V}, or with a redistribution of the stellar component due to secular radial migration \citep{Debattista06, Roskar08, Munoz-Mateos13}. The former view is supported by observations of U-shapes in the colour profiles of Type~II discs \citep{Bakos08, Chamba22} and by the removal of gas in dense environments that suppresses star formation in the outskirts of galaxies \citep{2022MNRAS.510.1716R}. In contrast, the latter scenario is supported by correlations between the break position and morphological features, such as bars and lenses, that can redistribute stars \citep{Laine14, Kim16, Laine16}. 

Type~III disc breaks can originate from a wide variety of physical processes, reflecting the diversity of disc galaxy evolutionary pathways. External mechanisms such as major or minor mergers \citep{2001MNRAS.324..685L, Borlaff14}, interactions with dark matter substructure \citep{Kazantzidis09}, and galaxy harassment in dense environments \citep{Roediger12} have all been proposed as formation channels. Tidally induced asymmetries may also be involved, particularly in S0 galaxies, which exhibit Type~III breaks more frequently than later-type ones, regardless of environment \citep{Erwin05, Gutierrez11, Maltby15}, suggesting a potential link between up-bending discs and the morphological transformation of spiral galaxies into lenticulars \citep{Borlaff14}. Additionally, some Type~III breaks represent transitions from discs to kinematically hot components such as stellar haloes or thick discs \citep{Pohlen06, Erwin08, Martin-Navarro12, Martin-Navarro14, Peters17, Comeron12, Comeron18}. Internal mechanisms are also relevant: stellar feedback and radial migration driven by bars or external accretion can steepen inner profiles or enhance outer discs, yielding anti-truncated profiles \citep{Hunter06, 2017MNRAS.470.4941H, Ruiz-Lara17, Clarke17, Okalidis22}. Observations indicate that outer discs beyond Type~III breaks may contain predominantly young populations, particularly in low-spin, gas-rich galaxies, supporting a connection with recent or ongoing star formation \citep{Laine16, Wang18}. Recent studies show a clear correlation between Type~III discs and their environments. Type~III discs are more commonly found in denser environments (\citealt{Watkins19}; hereafter \citetalias{Watkins19}). Conversely, they are significantly less frequent in isolated galaxies, where the few Type~III galaxies show distinct signs of merger activity (\citealt{Sanchez-Alarcon23}; hereafter \citetalias{Sanchez-Alarcon23}).

Due to the angular size, area, and depth limitations of earlier imaging surveys, statistical studies of breaks in disc galaxies have so far been limited to the nearby Universe ($z<0.1$). Only a few studies have examined disc breaks beyond the local Universe. For example, \cite{Azzollini08} found that the size of the disc, as defined by the break radius, decreases with increasing redshift. However, their limited sample (435 galaxies) and broad redshift coverage ($0.1<z<1.1$) prevented them from studying any evolutionary trend in the frequency of disc breaks. Similar results, showing that the disc size grows with time, were obtained using truncations as a tracer of size in a sample of 1000 galaxies observed by the \textit{Hubble} Space Telescope (HST) from $0<z<1$ \citep{Buitrago24}. While truncations and disc breaks are sometimes confused in the literature, they are structurally distinct features. Unlike disc breaks, truncations are defined by a sharp, rapid drop in the radial surface brightness profile at fainter surface brightness levels.
More recently, using the \textit{James Webb} Space Telescope (JWST), \cite{Yu25} confirmed the presence of disc breaks and U-shaped profiles in disc galaxies at $z\sim2$.

The \Euclid space mission \citep{Euclid} is set to revolutionise the study of galaxy structure and evolution, combining wide-area coverage with high spatial resolution and deep imaging sensitivity. Designed to survey an effective area of approximately 14\,000\,deg$^2$, \Euclid will observe a volume of the Universe orders of magnitude larger than previous space-based observatories such as the HST and the JWST, yielding an unprecedented number of resolved galaxies. Equipped with the visible imager (VIS), operating in a broad optical band \citep[\IE;][]{EuclidSkyVIS,Q1-TP002}, and the near-infrared spectrometer and photometer (NISP), which provides both photometry in three filters (\YE, \JE, \HE) and low-resolution spectroscopy \citep{EuclidSkyNISP, EuclidSkyNISPCU, Q1-TP003}, \Euclid is optimally configured to deliver high-fidelity multi-wavelength data across time. The instruments of \Euclid combine exquisite resolution [with pixel scales of $0\farcs1$ in VIS and $0\farcs3$ in NISP, and a point-spread function (PSF) full width at half maximum (FWHM) of $0\farcs16$ and $0\farcs48$, respectively] with remarkable surface brightness sensitivities, expected to reach down to $\sim$29.8\,mag\,arcsec$^{-2}$ in the Euclid Wide Survey (EWS) and $\sim$31.8\,mag\,arcsec$^{-2}$ in the Euclid Deep Survey \citep[EDS;][]{Scaramella-EP1, Borlaff-EP16} in the \IE band. These depths are defined as corresponding to 1$\sigma$ over $10^{\prime\prime}\times10^{\prime\prime}$ boxes. Such capabilities enable the detection and morphological characterisation of faint and extended galaxy features. The recent Q1 dataset, covering 63.1\,deg$^2$ with high-quality imaging \citep{EP-Aussel}, already surpasses the combined footprint of previous flagship extragalactic surveys, such as the $\sim$2\,deg$^2$ COSMOS field observed with HST \citep{COSMOS}, demonstrating \Euclid's potential to revolutionise statistical studies of galaxy morphology across large cosmic volumes.

Another of the major strengths of \Euclid lies in its unprecedented capability to detect low surface brightness (LSB) structures over cosmological volumes. The combination of exquisite image quality, stable PSF and space-based background, and uniform all-sky coverage  \cite[][]{EROData,Q1-TP004} allows \Euclid to probe diffuse stellar components, such as tidal features, stellar haloes, and outer disc structures, down to surface brightness levels previously accessible only in targeted deep fields. Within the Euclid Consortium, substantial effort is being devoted to the detection and characterisation of these faint structures, through both dedicated pipelines and coordinated working groups, ensuring that the mission fully exploits its potential for unveiling the LSB Universe.
 
In this work, we use the advantages of the \Euclid first quick release to investigate the disc breaks at different cosmic epochs. In Sect.~\ref{sec:EUC_Sample} we describe the sample selection criteria to select disc galaxies. In Sect.~\ref{sec:EUC_Pipeline}, we present the automatic pipeline developed for the photometric analysis, the creation of masks, the sky background estimation, the surface brightness extraction, and the disc break classification. In Sect.~\ref{sec:EUC_Results}, we review the results of the pipeline, assess its performance, and present the classification frequencies both overall statistics and across different redshifts. In Sect.~\ref{sec:EUC_discussion}, we discuss the results obtained. Finally, in Sect.~\ref{sec:EUC_conclusions}, we summarise the work and present our conclusions. 
Throughout, we use the AB photometric system, and we assume the $\Lambda$CDM model with a Hubble--Lemaître constant $H_0=70\,$km\,s$^{-1}\,$Mpc$^{-1}$, and a matter density parameter $\Omega_{\mathrm{m}}=0.3$.

\section{\label{sec:EUC_Sample}Sample selection}

This work is based on data from Q1 \citep[Q1;][]{Q1cite}, which comprises observations of the Euclid Deep Fields (EDF), covering a total area of 63.1\,deg$^2$. Each field was tiled with square footprints of 0.57\,deg$^2$ ($0\fdg75 \times 0\fdg75$), observed in a dithered pattern designed to optimise image quality and fill detector gaps. The total exposure times for each tile were approximately 2300\,s for VIS and 448\,s per NIR band, divided into four single exposures of 570\,s and 112\,s, respectively. Between exposures, the telescope performed a dither step of roughly 105$^{\prime\prime}$.
We refer the reader to \cite{Scaramella-EP1}, \cite{Q1-TP002}, and \cite{Q1-TP003} for a detailed explanation about the observation strategy and the reduction processing of VIS and NIR imaging. 

Furthermore, Q1 also provides external optical ground-based imaging. In the northern hemisphere, the EDF-N field has $ugriz$ optical imaging \citep[UNIONS;][]{Gwyn25}, while in the southern hemisphere, complementary $griz$ imaging is provided by the Dark Energy Survey \citep[DES;][]{DES,DES-DR2}, along with additional data from the DECam instrument on the Blanco Telescope \citep{DECAM}. A comprehensive description of the \textit{Euclid} mission design and survey strategy can be found in \citet{EuclidSkyOverview} and the details on the external optical data products in \citet{Q1-TP004}.

\begin{table}[t!]
    \centering
    \caption{Sample selection criteria.}
    \begin{tabular}{p{3.5cm}p{1.1cm}p{1.1cm}r}
        \hline \hline \\[-8pt]
        Parameter & Label & Criteria & Number    \\
        \hline \hline \\[-8pt]
        \texttt{smooth-or-featured \_featured-or-disk \_fraction}& $P_{\rm disc}$&$>0.6$ & 84\,071 \\
        \texttt{kron\_radius} & $R_{\rm Kron}$ & $>11^{\prime\prime}$ & 14\,667\\
        \texttt{phz\_pp\_median \_redshift} & $z$ & $<1$ & 11\,687\\
        |\texttt{phz\_pp\_median \_redshift $-$phz\_pp\_mode\_redshift}|\,/\,$z_{\rm mean}$ & $\Delta z$& $<0.01$ & 8748 \\
        \hline
    \end{tabular}
    \label{tab:euc_selection_parameters}
\end{table}

The Q1 data products were generated using the OU-MER pipeline \citep{Q1-TP004}, which delivers homogenised mosaic images, resampled to the same pixel scale of $0\farcs1$ for all the bands from \Euclid and external ground-based imaging, all centred at the same coordinates. The Q1 footprint is divided into 352 tiles with a 40\% overlap between tiles, with an angular size of $0\fdg53\times0\fdg53$, resulting in over 2908 images, consisting of 4\,TB of science-ready mosaics. Moreover, MER provides catalogues and photometry from VIS, NIR, and external data containing a variety of morphological parameters (see \citealt{Bertin20, Q1-SP040}, and \citealt{Q1-SP047} for more details) and photometric redshifts \citep[][]{Q1-TP005}. 

We aim to select a sample of disc galaxies, excluding earlier-type galaxies, to study their disc-type evolution through redshift. We base our selection of galaxies on the deep-learning morphological catalogue of \cite{Q1-SP047}. They selected galaxies with a segmentation area larger than 1200 pixels or with a $\IE<20.5$ and an area larger than 200 pixels. From the 370\,000 galaxies with visual morphologies in the version v4 catalogue \citep{Q1-SP047}, we select galaxies with a high probability of having a disc, with a disc probability $P_{\rm disc}$ greater than 60\%.  
To conduct a structural analysis of the stellar light distribution within these galaxies, we set a minimum Kron radius of 110 pixels ($11^{\prime\prime}$) for our sample ($R_{\rm Kron}>110\,$pixels). This value, while somewhat arbitrary, helps ensure both a high angular resolution, with a size much larger than the PSF of \Euclid \citep[][]{Q1-TP004}, and a statistically meaningful sample of galaxies. 
Additionally, we select galaxies with redshifts $z$ below one, and we require that the difference between the median and mode redshifts of the selected galaxies be less than 5\% of the mean of the mean and mode values. This criterion serves as a proxy for ensuring confident photometric redshifts ($ \Delta z<0.01$).

\begin{figure}[t]
    \centering
    \includegraphics[width=\linewidth]{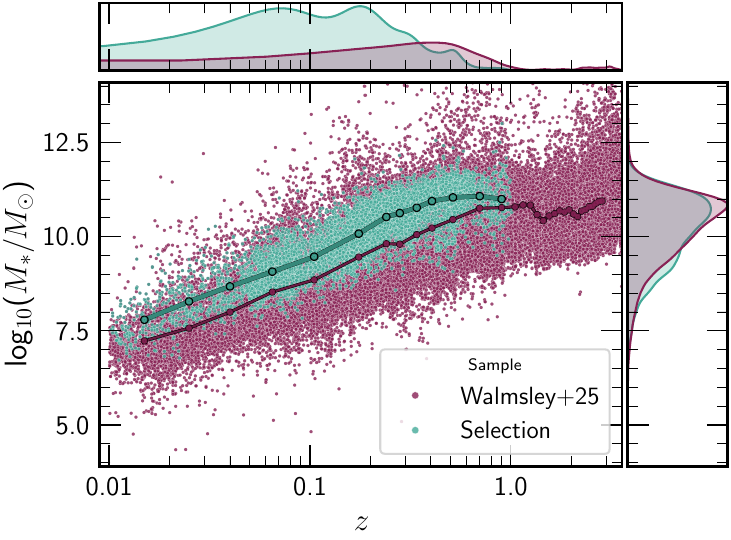}
    \caption{Stellar mass as a function of redshift for the galaxies in the different samples. The stellar mass and redshift are obtained by the \texttt{PHZ} processing function. We show the master sample, from which the selected sample is drawn, in violet, while the final sample is shown in blue. The \textit{middle} panel shows the stellar mass with respect to the redshift of the galaxies. The side plots show the density distribution of the redshift (\textit{top}) and the stellar mass (\textit{right}).}
    \label{fig:sample}
\end{figure}

This selection criterion results in a sample of 8748 sufficiently resolved disc galaxies with accurate redshifts. Table~\ref{tab:euc_selection_parameters} shows the column names of the different catalogues, their label used in the text, their selection conditions, and the resulting sample size after each criterion has been applied. Figure~\ref{fig:sample} shows the stellar mass with respect to redshift from the master samples from \cite{Q1-SP047} and our selection. At lower redshifts, we are biased towards less massive galaxies, since the limited area of Q1 results in a lack of massive galaxies ($M_*\gtrsim10^{9.5}\,M_\odot$) at low redshift ($z<0.06$). This effect will be reduced with future releases of the EWS.

 \section{\label{sec:EUC_Pipeline} Surface brightness profile analysis pipeline}
%\pablo{Use past tense} \\

\begin{figure*}
    \centering
    \includegraphics[width=0.98\linewidth]{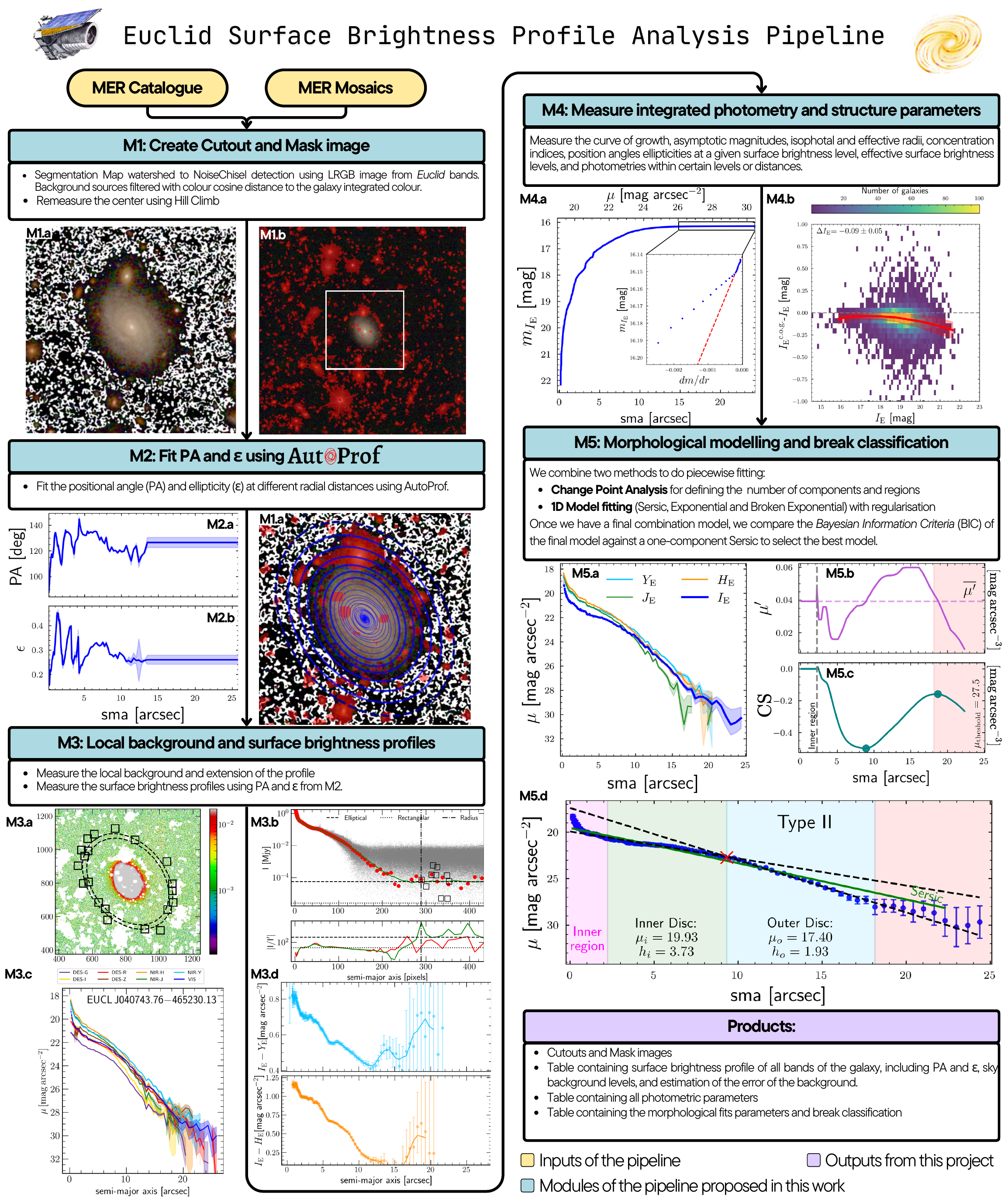}
    \caption{Figure illustrating each module (M) in the automatic pipeline to analyse the galaxy EUCL\,J040743.76$-$465230.1. Each module is represented by a header (blue boxes), with a short description. More details of the modules can be found in Sect.~\ref{sec:EUC_Pipeline}. The figures below each box are associated with the methods of each module. In M1, a colour image of the galaxy (\textit{left}) and the same image with the mask overlayed in red (\textit{right}) are shown. In M2, the resulting PA and $\epsilon$ profiles (\textit{left}) and the resulting elliptical apertures on top of the image (\textit{right}) are shown. The top figures of M3 show the rectangular and elliptical apertures used to measure the statistics of the background (\textit{left}), and the $I_{I_\sfont{E}}$ and $I_{I_\sfont{E}}/I_{I_\sfont{E}}^\prime$ profiles used to determine. the position of the apertures (Sect.~\ref{sec:EUC_M3_bkg}). The \textit{bottom} figures show the resulting profiles (\textit{bottom left}) and the colours from the \Euclid bands (\textit{bottom right}). The M4 panel shows the curve of growth (\textit{left}), and the comparison between \IE magnitudes between our method and MER (\textit{right}). The M5 panel shows the \IE profile ($\mu_{I_\sfont{E}}$, \textit{top left}) and its derivative ($\mu_{I_\sfont{E}}^\prime$) with the cumulative sum used in the change point analysis. The resulting fitted models, given the regions defined, are shown in the \textit{bottom} part of M5 (Sect.~\ref{sec:EUC_M5} for more details).}
    \label{fig:EUC_pipeline}
\end{figure*}

For each galaxy in the sample, we run an automatic pipeline designed to extract surface brightness profiles and structural and integrated quantities, and classify breaks according to their different type of discs (Type~I, Type~II, Type~III, and a combination of Type~II and Type~III breaks). The entire pipeline is subdivided into five different modules. Module~1 (M1) produces cutouts and mask images, while Module~2 (M2) measures position angle and ellipticity radial profiles. The sky background value, the extent of the profile, and the surface brightness profiles on the different bands are measured in Module~3 (M3). Photometric quantities are measured in Module~4 (M4). Finally, Module~5 (M5) classifies profiles into the three different types of disc breaks. In the following subsections, we explain in detail the different methods used in each module. 
We show an illustration of the pipeline and each of the modules in Fig.~\ref{fig:EUC_pipeline} which show an example of the analysis of the galaxy EUCL\,J040743.76$-$465230.1. All figures are explained further in each subsection.\\

\subsection{\label{sec:EUC_M1}Module 1: Cutouts and masks images}

We construct cutouts for each band available for each galaxy centred on the coordinates of the source and with a size of five times the Kron radius ($5\, R_{\rm Kron}$). The resulting large cutouts allow us to study the sky background and produce accurate masking. For each of the cutouts, we add the background model resulting from MER, as this step can mitigate any potential sky background oversubtraction, as is known to occur in some galaxies \citep{EP-Urbano}. To refine the central coordinates and account for any misalignments, we remeasure the centre using the \software{Hill Climb} method of \software{AutoProf} \citep{Autoprof}. 
\software{AutoProf} employs an iterative algorithm that uses FFT-based gradient detection, parabolic fitting, and Nelder--Mead optimization to converge on the galaxy's global flux peak. This method effectively ignores local features to determine the primary center of the galaxy.

Masks are a crucial resource for producing reliable surface brightness profiles. We combine the best software and techniques to create the masks. We start with the segmented image from MER, which is a product of using \software{SourceXtractor++} on the mosaics and additional methods to perform deblending \citep[details in][]{Q1-TP004}. We run \software{NoiseChisel} \citep{noisechisel} on the \IE-band image to detect all pixels with signal. \software{NoiseChisel} has been demonstrated to be optimal at detecting signal down to low signal-to-noise ratios (S/N) levels \citep[][]{noisechisel}. We create an LRGB image using channels \YE, \JE, and \HE, for colours and \IE for a luminance mask, stretched using the asinh transformation. We then grow the segments of the MER image to fill the detection map of \software{NoiseChisel} using the watershed algorithm (from the \texttt{scikit-image.segmentation} library; \citealt{scikit-image}; v0.24.0) on the LRGB image. While we could have used either of the bands provided by \Euclid, the colour information from the LRGB images yielded better results for deblending sources after testing.

We use \software{MTObjects} \citep{mtobjects} on the \IE image to deblend sources inside the galaxy of interest with a \software{move\_factor} set to 0.3. However, to filter external sources from real regions of the galaxy, we use the cosine similarity $\vec{S_C}$ between colours produced from the \JE, \YE, \JE, and \HE band images. We do not use the conventional astronomical definition of colour (i.e. the magnitude difference between two bands), but instead refer to the multi-band intensity vector defined by the fluxes measured in the different photometric bands. The cosine similarity is a distance measurement often used in computer vision to quantify the difference in direction between two vectors, specifically the angle between them. We measure the cosine similarity between the vector defined by the colour of the galaxy and the colour vector of each pixel. We first measure the average intensity of the galaxy, over the segmentation area in the \Euclid detection mask, for each \Euclid band, and we define the colour vector $\vec{g}:=(g_{I_\sfont{E}},g_{Y_\sfont{E}},g_{J_\sfont{E}},g_{H_\sfont{E}})$. For a pixel at spatial coordinates $i$ and $j$, referenced by $C^{ij}_b$, in the colour image $\vec{C}$, we measure the cosine distance to the galaxy as:
\begin{equation}
\vec{S_C} = \frac{\vec{C} \cdot \vec{g}}{\| \vec{C} \| \, \| \vec{g} \|}=\frac{\sum_{b} C_b g_b}{\sqrt{\sum_{b} C_b^2} \, \sqrt{\sum_{b} g_b^2}} \; ,
\end{equation}
where $b$ runs over the \Euclid bands in the predefined order.
The cosine similarity image $\vec{S_C}$ is an estimate of the colour similarity of each pixel in the image $\vec{C}$ to the integrated colour of the galaxy, $\vec{g}$, using \Euclid's bands. For each region deblended by \software{MTObjects}, we measure the median of the cosine similarity. If the difference between the regions and the one from the galaxy is larger than three times the standard deviation $\sigma_{\vec{g}}$ of the pixels inside the region of the galaxy, we assume it has a different colour, thus it is an external source, and we mask it. If the difference is lower than $3\,\sigma_{\vec{g}}$, we assign the same ID to the source as to the galaxy of interest. 

\begin{figure*}[t]
    \centering
    \includegraphics[width=\linewidth]{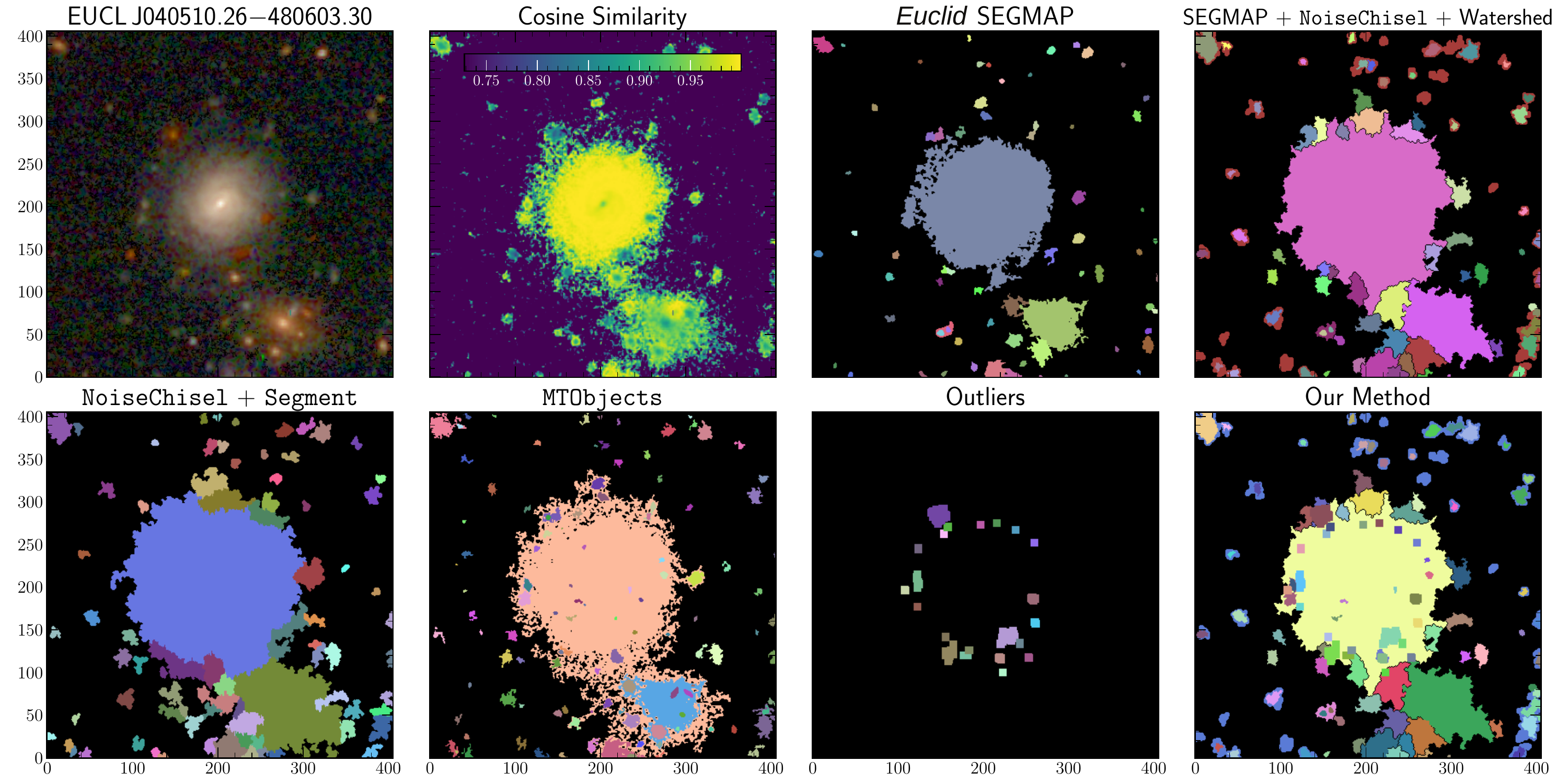}
    \caption{Comparison of masks for EUCL\,J040510.27$-$444836.3. In the \textit{top} row, we display from \textit{left} to \textit{right}, the LRGB image of the galaxy, the cosine similarity map $\vec{S_C}$, the MER segmentation map, and the resulting segmented image from the watershed process. In the \textit{bottom} row, from \textit{left} to \textit{right}, we show the resulting segmentation map from running  \software{NoiseChisel+Segment}, \software{MTObjects}, and the regions found using the cosine similarity map and the resulting segmentation from our method (Sect.~\ref{sec:EUC_M1} for more details).}
    \label{fig:segmentation}
\end{figure*}

The last step in producing the mask images is to search for outliers in the colour space inside the galaxy region. We select pixel values that are larger than $5\,\sigma_{\vec{g}}$. These regions correspond typically to high-redshift sources, which are not deblended by \software{MTObjects} due to the smoothness and similarity in brightness on the \IE channel. To remove spurious detections (individual pixels with large values), we erode and dilate the mask produced by using a $5\,\sigma_{\vec{g}}$ threshold. Then, all sources with an area larger than the FWHM$^2$ are masked. This extra step is useful for deblending sources with very different colours (which are probably not part of the galaxy) and extended, faint emission, such as red, point-like background sources. In all these steps, to prevent masking the central region of the galaxy, we left unmask the region within 20\% of the size of the region in the mask corresponding to the galaxy. 

Figure~\ref{fig:segmentation} shows the resulting mask from the method explained applied to a galaxy of the sample, 
EUCL\,J040510.27$-$444836.3. We also show the segmentation results from using standard algorithms, such as \software{NoiseChisel+Segment} \citep{noisechisel} and \software{MTObjects}, and the \Euclid segmentation map, for comparison. We show a few intermediate steps such as the cosine similarity map, the watershed resulting mask, and the outliers identified using \software{MTObjects} and the cosine similarity map. In the top panels of M2 in Fig.~\ref{fig:EUC_pipeline}, we show a cutout example (colour image on the left) and the resulting mask (right), showing the mask using red regions for EUCL\,J040743.76$-$465230.1.

\subsection{\label{sec:EUC_M2}Module 2: Position angle and ellipticity radial profiles}
We measure the position angle PA and ellipticity $\epsilon$ radial profiles using the new implementation of the \cite{Jedrzejewski} method on \software{AutoProf} by \citet{Autoprof}. We use the function \software{Isophote\_Fit\_FFT\_Robust} of \software{AutoProf} that fits elliptical apertures to the galaxy image for a given set of radial distances. 
For each isophotal fitting step, the function uses a method similar to that in \cite{Jedrzejewski} with a few regularisation terms in the loss function that improves the fitted results and speed. We use an increment in the radial distance of the profiles by a logarithmic rate, setting the consecutive radial steps to be located at positions such as $r_i = 1.03\,r_{i+1}$. 
This logarithmic difference allow us to finely trace the inner regions while increasing the apertures in the outskirts of the galaxy with lower S/N. We use the mask images from Sect.~\ref{sec:EUC_M1} to remove contaminating light from other sources while fitting the apertures. 
We fit the PA and $\epsilon$ parameters of each elliptical aperture until we reach levels of 3\% of the noise measured in each image (setting the parameter \software{ap\_fit\_limit} to 0.03). The regularisation parameter (\software{ap\_regularize\_scale}) is set to 1.5 to penalise hard jumps between radial isophotes (see \citealt{Autoprof} for more details). We estimate the noise as the sigma-clipped standard deviation from the pixels not masked in the M1 mask image to give it to \software{AutoProf}.
In panel M2 (left middle) in Fig.~\ref{fig:EUC_pipeline}, we show the resulting PA (upper left panel) and  $\epsilon$ (lower left panel) profiles, and we show the elliptical apertures on top of the colour image with the mask in red (right panel).  

\subsection{\label{sec:EUC_M3}Module 3: Sky background and surface brightness profiles}

For all the galaxies in the sample, we use the radial PA and $\epsilon$ profiles to measure the local sky background around the galaxy and set the extent of the profiles. Then, we measure the surface brightness profiles on all the available bands. We describe the details of each procedure below.  

\subsubsection{\label{sec:EUC_M3_bkg}Sky background and profile extension}

We measure a fixed elliptical aperture intensity profile of the galaxy by setting the PA and $\epsilon$ to be the average of the outermost five values of the PA and $\epsilon$ \software{AutoProf} profile. These PA and  $\epsilon$ values follow the morphology of the outskirts of the galaxy. We measure the intensity profile, $I_b$,  with elliptical apertures at each radial distance, $I_b(r)$, until we reach the end of the image, for each \Euclid band, $b$. 
The radial surface brightness profile of a galaxy is often well described by an exponential profile, either a Sérsic or a pure exponential profile. We aim to find the end of these components and find the radius where the background dominates. This transition is usually seen as a flattened region in the intensity profile (or a minimum followed by a plateau where there are oversubtraction artefacts). If we assume that all galaxies have a pure exponential profile in their outskirts, we can find this transition by studying the parameter $I_b/I^\prime_b$, where $I^\prime_b$ is the derivative with respect the radius, $\mathrm{d} I_b/\mathrm{d}r$. We measure the derivative by fitting a slope to the five adjacent points in the profile in counts \citepalias[similar to][]{Watkins19, Sanchez-Alarcon23}. For a pure exponential profile, the parameter $I_b/I^\prime_b$ is constant, while for a plateau (i.e., an horizontal line), the derivative of the intensity profile tends to zero ($I_b^\prime \rightarrow0$), hence, $I_b/I^\prime_b$ tends to infinity. Then, the $I_b/I^\prime_b$ profile has a peak at the radius where the transition happens. We set the maximum radius of the profile $R_{\rm max}$ to be equal to the radius where the plateau is found. The plateau was reached when $I_b/I^\prime_b$ deviate more than three times the standard deviation of the average value. We smooth the intensity profile $I_b$ using the implementation of the Savitzky--Golay filter \citep{SG-Filter} by \software{scipy} (\citealt{scipy}; v1.15.2) with a window of five and a polynomial degree of one. By smoothing the profile, we remove the noise due to local asymmetries and contamination light that could introduce artefacts on the $I_b/I^\prime_b$ profile.

We set the background value to be the sigma-clipped median of the non-masked pixels inside an elliptical annulus at $R_{\rm max}$ with a width of $0.1\, R_{\rm max}$. Using square boxes at a distance of $R_{\rm max}$, we measure the uncertainty of the background value as the standard deviation of the medians of the pixels inside the rectangular boxes, normalised by the square root of the number of boxes. The number of boxes and the width are set by the distance $R_{\rm max}$.

The panels M3 in Fig.~\ref{fig:EUC_pipeline} show an example of the procedure of background and size estimation. We show the \IE image of  EUCL\,J040743.76$-$465230.1 (top left), setting the minimum and maximum values of the visualisation around the background value. The top right panel shows the intensity profile ($I_b$) measured using fixed elliptical apertures. The grey dots represent all the pixels in the image, the red dots are the sigma-clipped average of the pixels within the isophotes, while the smoothed profile using the Savitzky--Golay filter is in green. In a lower panel, we also show the $I_b/I^\prime_b$ profile for the measured (red curve) and smoothed (green curve) profiles. The $I_b/I^\prime_b$ profile is nearly constant out to $\sim$250 pixels, where the $I_b$ profile reaches a plateau. At around $\sim$300 pixels, the $I_b/I^\prime_b$ profile is above the threshold, which is defined as the 5$\sigma$ distance from the mean of the entire $I_b/I^\prime_b$ profile. The $R_{\rm max}$ value is set to this location, and the background is measured based on this radial distance. The squares in the profile represent the mean values of each sky box.

\subsubsection{\label{sec:EUC_M3_prof}Surface brightness profiles}
Once we set the extension of the profile, we measure the surface brightness profiles for each band available. We use elliptical regions that followed the PA and $\epsilon$ \software{AutoProf} profile until $R_{\rm max}$. We measure the surface brightness using sigma-clipping averages of the non-masked pixels inside the elliptical aperture. We also measure the flux inside the aperture. To overcome the flux lost in masked pixels, we assigne the average surface brightness within the aperture to those masked pixels, which we add to the total flux of the aperture. 
    
The lower panels of  M3 in Fig.~\ref{fig:EUC_pipeline} show the profiles measured using the elliptical apertures from M2 up to $R_{\rm max}$ for all the bands available for this galaxy (on the left), and the \Euclid colour profiles (on the right). 

\subsection{\label{sec:EUC_M4}Module 4: Derived parameters}
Using the derived profiles, we measure integrated magnitudes and morphological parameters (PA and $\epsilon$) at different isophotal levels, isophotal radii, and concentration indices. We measure the same parameters as in the \textit{Spitzer} Survey of Stellar Structure in Galaxies (S$^4$G; \citealt{Seth10,Munoz-Mateos13, Watkins22}) following the methods used for the Complete \textit{Spitzer} Survey of Stellar Structure in Galaxies (CS$^4$G; \citealt{Sanchez-Alarcon25}; hereafter \citetalias{Sanchez-Alarcon25}). More details on the parameters can be found in their section~6. 

In a nutshell, we derive asymptotic magnitudes as explained by \cite{Watkins22}, by measuring the $y$-intercept of the linear fit on the final part of the flux profile (an example of the process is illustrated in the left panel of M4 in Fig.~\ref{fig:EUC_pipeline}). Moreover, we derive different estimates of the galaxy radius such as the half-light or effective radius $R_{\rm e}$, and the isophotal radii at levels of 25.5, 26.5, and 28 mag\,arcsec$^{-2}$ ($R_{25.5}$, $R_{26.5}$, and $R_{28}$, respectively). We also measure the PA and $\epsilon$ and the different surface brightness levels, as well as two concentration indices, $C_{82}$ and $C_{31}$.

The right panel of M4 in Fig.~\ref{fig:EUC_pipeline} shows the difference between the MER magnitudes and our asymptotic magnitudes for \IE. We find good agreement with the values reported by MER, and we recover 0.09\,mag on average for the sources in our sample using our method. A detailed comparison with MER can be found in Appendix~\ref{app:EUC_MER}.

\subsection{\label{sec:EUC_M5}Module 5: Disc modelling in 1D}
 
We infer the disc break classification using a piecewise fit of the profile by combining two methods. In summary, we first divide the profile into different regions using a change-point analysis \citepalias[similar to][]{Watkins19,Sanchez-Alarcon23}, and second, we fitted models to each region to define a combined model that describes the galaxy profile. Finally, we test if the combined model is a better fit than a simple Sérsic profile fit. 

We now explain this approach in detail. We apply this technique to the surface brightness profiles by transforming the intensity profile $I_{I_\sfont{E}}$ using the zero point of each image, and its pixel scale ($0\farcs1$), to  mag\,arcsec$^{-2}$ as
\begin{equation}
    \mu_{I_\sfont{E}} = {\rm ZP}_{I_{\rm E}}- 2.5\log_{10}\left(\frac{I_{I_\sfont{E}}}{1\,{\rm ADU/s}}\right) + 5\log_{10}\left(\frac{\rm scale}{1\,{\rm arcsec}}\right) + c \;,
\end{equation}
where $\mu_{I_{\rm E}}$ is in mag\,arcsec$^{-2}$, ZP is the zeropoint of the image, scale refers to the pixel scale of the \IE band in arcsecs, and $c$ corresponds to some corrections applied to the profile. The corrections are:
\begin{itemize}
    \item Redshift dimming proportional to $(1+z)^3$, as derived by \cite{Ribeiro2016}:
    \begin{equation}
    c_1 = -7.5\log_{10}(1+z) \;.
    \end{equation}
    \item Inclination correction, following \citetalias{Sanchez-Alarcon25}:
    \begin{equation}
     c_2 = -2.5\log_{10}(1-\epsilon) \;,
    \end{equation}
    where $\epsilon=b/a$ is the axis ratio, the ratio between the semi-minor ($b$) and semi-major ($a$) axes. We measure $\epsilon$ using the ellipticity profile and taking the the median of values with surface brightness fainter than $25.5\,$mag arcsec$^{-2}$. 
    %\item Extinction correction from the value of the MER?
\end{itemize}
The total correction $c$ is the sum of the individual components $c=c_1+c_2$.
We do not correct for Galactic extinction as it is not available for \Euclid bands.
We apply a surface brightness threshold of $27.5\,$mag\,arcsec$^{-2}$ in the rest frame of the galaxy to ensure that we observe the same surface brightness range across the entire redshift coverage. We trim all the values fainter than this threshold before analysing the disc components. This threshold was set to reach the approximate depth limit of \Euclid, i.e. $29.5\,$mag\,arcsec$^{-2}$, at $z\sim1$, i.e., $27.5\,$mag\,arcsec$^{-2}$ restframe. 

With the surface brightness profile corrected and the threshold applied, we separate the profile into different regions, and determine the peaks of the cumulative sum of the profile derivative minus the mean of the derivative. We follow the approach used in \citetalias{Watkins19} and \citetalias{Sanchez-Alarcon23}. This method, usually referred to as change-point analysis, detects changes in the derivative of the profile, which reflect the changes in the slope of the exponential disc. We measure the derivative $\mu^\prime$ of the profile by fitting a line using the five adjacent points (as in M3, see Sect.~\ref{sec:EUC_M3_bkg}). 
This way we smooth the $\mu^\prime$ while preserving the global slope and removing high-frequency noise. We define the mean ($\bar{\mu^\prime}$) of $\mu^\prime$ as the weighted mean using as weights the distance in arcsec between points to account for the logarithmic growth between points in the profiles. We define the cumulative sum as 
\begin{equation}
    {\rm CS}_0=0, \quad {\rm CS}_N=  \sum_{j=1}^N\mu^\prime(j)-\bar{\mu^\prime}\ \;.
\end{equation}
We search for peaks in CS with a prominence higher than $0.15\,$mag\,arcsec$^{-3}$. Using the number of peaks detected in the cumulative sum, $N_{\rm peaks}$, we define $N_{\rm peaks} + 1 $ regions in the profiles. We performed a piecewise fit for each region, using different models, including Sérsic, exponential, and broken exponential profiles, and select the best fit according to specific criteria based on their $\chi^2$ (see below). The M5 panels in Fig.~\ref{fig:EUC_pipeline} show as an example of the method, the derivative of the profile (top left) and the CS profile (bottom left). For this example, the CS only shows one minimum (after removing the inner region) within the surface brightness threshold; thus, the profile can be divided into up to two components. 

For the inner region, between the centre and the radius where the first peak occurs, we always fit a Sérsic profile. We use the Sérsic profile adapted to 1D and in units of mag\,arcsec$^{-2}$ \citep[Equation~6 of][]{2005PASA...22..118G}. We fit the parameters $\mu_{\rm e}$, $R_{\rm e}$, and $n$ by minimizing the $\chi^2$. \\

\begin{table*}[ht!]
    \centering
    \caption{Break type classification and statistical results from M5.}
    \label{tab:break_combined}
    \begin{tabular}{p{2.5cm}p{2.5cm}p{5.5cm}|>{\raggedleft\arraybackslash}p{2.5cm}>{\raggedleft\arraybackslash}p{2.5cm}}
        \hline \hline \\[-8pt]
        Type & \# of discs & Criteria & Count & Fraction (\%) \\
        \hline \hline \\[-8pt]
        0         & 0   & Only Sérsic, no disc                   & 120  & 3  \\
        I         & 1   & Only one disc                           & 719  & 17 \\
        II        & 2   & $h_1 > h_2$                           & 1518 & 35 \\
        III       & 2   & $h_1 < h_2$                           & 623  & 14 \\
        II+II     & 3   & $h_1 > h_2$ \software{}{AND} $h_2 > h_3$ & 48   & 1  \\
        III+III   & 3   & $h_1 < h_2$ \software{}{AND} $h_2 < h_3$ & 145  & 3  \\
        II+III    & 3   & $h_1 > h_2$ \software{}{AND} $h_2 < h_3$ & 874  & 20 \\
        III+II    & 3   & $h_1 < h_2$ \software{}{AND} $h_2 > h_3$ & 233  & 5  \\
        Null      & --  & Not classified                       & 105  & 2  \\
        \hline \\[-8pt]
         &  & \hspace{4.2cm} \textbf{Total:}     & \textbf{4385} & $\pm 1\%$ \\
        \hline
    \end{tabular}
    \tablefoot{The numbering of the scale lengths follows radial distance from the centre, starting with $h_1$.}
    \label{tab:break_class}
\end{table*}

Then, for the rest of the regions, we follow these steps:
\begin{enumerate}

    \item We fit a broken exponential. We use the same function as \cite{2016A&A...587A..70S}. We define the broken exponential as the sum of two linear functions truncated at the radius of the break $r_{\rm break}$. The function, in mag\,arcsec$^{-2}$, is defined as

    \begin{align}
    \label{eq:broken_exp}
    \mu(r) &=\left(\mu_{\rm in} + \frac{2.5}{\ln 10} \, \frac{r}{h_{\rm in}} \right) \,\left(1-W_{\rm b}\right)+ \nonumber\\
    &+\left(\mu_{\rm out} + \frac{2.5}{\ln 10} \, \frac{r}{h_{\rm out}} \right) \, W_{\rm b} \;,
    \end{align}
    
    where $\mu_{\rm in,out}$ is the surface brightness level at the centre, and $h_{\rm in,out}$ is the scale length of the disc. The in- and out-subindices denote the inner and outer discs. The parameter $W_{\rm b}$ is a softening parameter for the transition between the linear relations, 
    
    \begin{equation}
    W_{\rm b} =\frac{\pi / 2+\arctan \left(\frac{r-r_{\rm b}}{\beta}\right)}{\pi}\;,
    \end{equation}
    
    and we impose the continuity condition at the radius of the break $r_{\rm b}$
    
    \begin{equation}
    \mu_{\rm in} + \frac{2.5}{\ln 10}\,\frac{r_{\rm b}}{h_{\rm in}} = \mu_{\rm out} + \frac{2.5}{\ln 10}\,\frac{r_{\rm b}}{h_{\rm out}} \;.
    \end{equation}
    We set a value of $\beta = 10^{-8}\,$arcsec to have a sharp break, and we fit this function leaving the parameters $h_{\rm in}$, $\mu_{\rm in}$, $h_{\rm out}$, $\mu_{\rm out}$ free. The fitting process consists in minimising the loss function
    \begin{align}
        l &= \chi^2 + \frac{\alpha}{(h_{\rm in}-h_{\rm out})^2/h_0^2 + \delta} +  \nonumber\\
        &+ \left [ \frac{\gamma}{1+{\rm e}^{3\,(r_{\rm b}-r_{\rm min})/r_0}} + \frac{\gamma}{1+{\rm e}^{2\,(r_{\rm max}-r_{\rm b})/r_0}} \right ] \; ,
    \end{align}
    where $r_0 = h_0 = 1\,$arcsec is introduced to maintain the dimensionless properties of the equation. We add two regularisation terms to the $\chi^2$. The first one avoids finding a similar scale length in both discs, while the second one avoids finding breaks very close to the borders ($r_{\rm min}, r_{\rm max}$). We chose the values of $\alpha$, $\delta$, and $\gamma$ to be $10^{-2}$, $10^{-4}$, and the number of data points inside the region, respectively. \\  
    
    \item We fit a single exponential to the same region. We use the exponential function in surface brightness (first parenthesis in Eq.~\ref{eq:broken_exp}).  We minimise the $\chi^2$ to perform the fit. \\

    \item We select the model that best described the profile in this region. To select the broken exponential, the model has to meet the following  conditions:\\
    
    \begin{enumerate}
        \item The values of $h_{\rm in}$ and $h_{\rm out}$ differ by at least 10\% following \citet{Wang18} and \citet{Yu25}.
        
        \item The Bayesian information criteria (BIC) of the broken exponential model must are lower than those of the single exponential. 
    \end{enumerate}
    
    The BIC is a measurement of the $\chi^2$ that penalises complex models. We use
    \begin{equation}
        \mathrm{BIC}=N \ln \left(\chi^2 / N\right)+\ln (N) N_{\text{varys}}\; ,
    \end{equation}
    where $N$ is the number of data points being fitted, and $N_{\rm varys}$ is the number of variables of the model. 
\end{enumerate}

After completing these three steps for each region, we create a model that combines the resulting components, consisting of a maximum of one Sérsic component and up to four disc components. There is clear evidence from the literature of galaxies with two or even three disc components \citep[e.g.][\citetalias{Watkins19}]{Laine14}. We set a maximum of four disc components to avoid unnecessarily limiting the models. We impose the continuity condition between all discs components. With the combine model, we fit again the profile with all the parameters of the components, as well as the regions defining each parameter, minimising only the $\chi^2$. We set the initial parameters to the parameters fitted in the previous steps. We compare the BIC of the combined model with a single Sérsic fit to ensure that the profile is better described by a combination of discs than a single Sérsic profile. We select the model with the lowest BIC as the final one. 

It is then straightforward to classify the profiles according to their break types. We classify the profile according to the scale~lengths found and following the criteria shown in Table~\ref{tab:break_class}. We limit the classification to three types of disc, and profiles best described with four discs are classified as `Null'. 

The bottom panel of M5 in Fig.~\ref{fig:EUC_pipeline} shows an example of a surface brightness profile classified in M5. The resulting two components from the CS were fitted using a broken exponential, resulting in a Type~II profile that reproduces successfully the shape of the profile.  

\section{\label{sec:EUC_Results}Results}

Using the automatic pipeline developed and explained in this work, we analyse 8748 galaxies from our selected sample (Sect.~\ref{sec:EUC_Sample}). We execute the pipeline on the \texttt{diva} computer of the IAC using 90 Intel(R) Xeon(R) CPU E7-4850 v4 @ 2.10GHz cores with 270\,Gb of RAM. It took 18\,hours to process all the galaxies in the sample with an average CPU-efficiency of 98\%. 

The pipeline successfully ran for 7873 (90\%) of galaxies through M1 to M4. 
Appendix~\ref{app:EUC_MER} provides a comparison of our asymptotic integrated magnitudes and MER photometry.  
M5 successfully classified 6339 (73\%) galaxies\footnote{The fitting procedure did not converge for the remaining galaxies.} into the different disc types as outlined in Table~\ref{tab:break_class}. For the following analysis, we selected galaxies with reduced $\chi^2_\nu$ and with maximum values in the $\chi^2(r)$ radial profile ($\chi^2_{\rm max}$) by sigma-clipping their distribution, removing galaxies within three times the standard deviation, ensuring the selection of galaxies with good fits. This selection resulted in 4385 galaxies with $\chi^2_\nu < 0.75$\footnote{To calculate $\chi^2_\nu$ each profile has a particular degree of freedom d.o.f., but the average d.o.f is 40.} and $\chi^2_{\rm max}<7.4$, and hereafter, we will refer to it as the reliable sample. 

We present the results of the classification of galaxies, their structural parameters, and their evolution with redshift in the following subsections. 

\subsection{Disc break type classification}

In Table~\ref{tab:break_class}, we present the classification of disc-break types for the 4385 galaxies successfully model with high confidence. Since our fitting procedure does not impose any restriction on the number of components, we also identify composite profiles exhibiting multiple breaks, such as Type~II+III systems, where an inner down-bending (Type~II) break is follow by an outer up-bending (Type~III) break, among other combinations. The most frequent profile in our sample is Type~II, comprising 1518 galaxies (35\%), followed by Type~II+III with 874 (20\%), Type~I with 719 (17\%), and Type~III with 623 (14\%). The remaining 11\% correspond to less common composite types: Type~III+II with 233 (5\%), Type~III+III with 145 (3\%), Type~II+II with 48 (1\%), and 120 galaxies (3\%) best described by a single Sérsic profile. Only 105 galaxies (2\%) required more than four components and were therefore classified as \texttt{Null}.

To assess the reliability of the M5 classification procedure, we visually inspect a random subsample comprising more than 10\% (490 galaxies) of the confident sample. We assign to each galaxy a quality flag based on visual inspection of the images, masks, surface brightness profiles, and resulting fits:
\begin{enumerate}
    \item Correct: The automatic classification is consistent with the visual inspection and accurately captures the surface brightness profile features.
    \item Unsure: The classification is reasonable, but the surface brightness profile shows ambiguities or smooth transitions between components that could suggest alternative interpretations. An example could be a smooth transition from an inner component (Sérsic) to an outer disc that could be interpreted as a Type~II profile, but it is not detected nor clear in the visual inspection.
    \item Incorrect: The classification fails to reproduce the observed surface brightness features and does not accurately model the disc structure.
    \item Problematic: The profile is deemed unreliable due to external factors such as misclassification of a non-galaxy object, centring issues, or poor masking. 
\end{enumerate}

Of the 490 visually classified galaxies, we label 344 (70\%) as `correct', 59 (12\%) as `unsure', 73 (15\%) as `incorrect', and 14 (3\%) as `problematic'. We therefore adopt an overall classification accuracy of approximately 70\% for M5.

\begin{figure*}[ht!]
    \centering
    \includegraphics[width=0.9\linewidth]{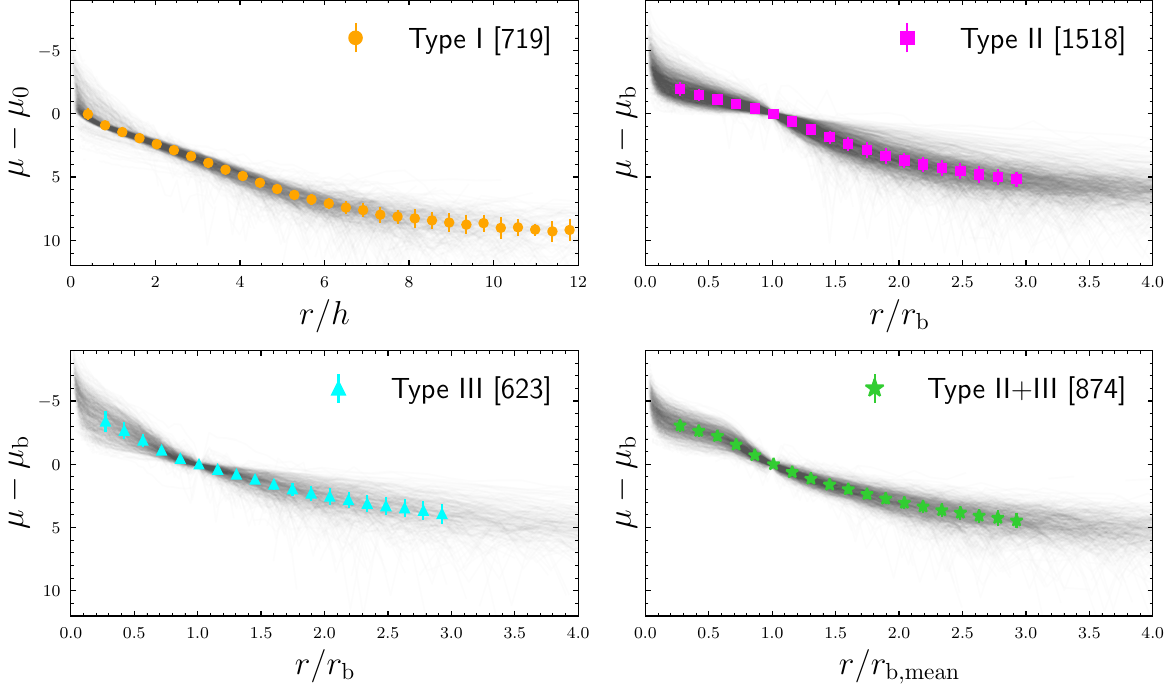}
    \caption{Stacked profiles of galaxies grouped by disc break type, using the classification obtained from M5. The panels, from \textit{top-left} to \textit{bottom-right}, include all the profiles (text and Table~\ref{tab:break_class}) from our analysis categorised as Type~I, II, III, and II+III. 
    The surface brightness ($\mu$ in mag\,arcsec$^{-2}$) is scaled to the brightness at the break point $\mu_{\rm b}$ or, for Type~I profiles, at the central surface brightness $\mu_0$, while the radius is normalised to the radius of the break $r_{\rm b}$ or to the scale length $h$ for Type~I profiles. For Type~II+III galaxies, the profiles are scaled according to the mean values of both breaks (see text).
    The grey curves represent the profiles of all galaxies, and the dots indicate the median of all profiles. 
    The legends specify the number of profiles (in brackets) stacked in each category. Error bars are estimated by measuring the root mean square of all values within each radial bin.}
    \label{fig:EUC_types}
\end{figure*}

Figure~\ref{fig:EUC_types} shows the stacked, normalised surface brightness profiles for the most common profile types, Type~I, II, III, and II+III. Profiles are normalised to the break radius and the surface brightness at the break, and for Type~I galaxies, profiles are scaled using the scale~length and central surface brightness of the disc. For Type~II+III profiles, we use the mean radius $r_{\rm b,mean}$ and surface brightness $\mu_{\rm b,mean}$ of the inner and outer breaks for normalisation. The colour markers indicate the running mean of the profiles. In the top-left panel, the average Type~I profile reveals a pure exponential disc extending out to $\sim$8 scale~lengths. The Type~II and III profiles exhibit down-bending and up-bending breaks, respectively. In Type~II+III, the Type~II breaks typically occurs at $r/r_{\rm b,mean} \sim 0.8$, and the Type~III occurs at $r/r_{\rm b,mean} \sim 1.3$.

\subsection{\label{sec:EUC} Structural distribution of disc types parameters}

\begin{figure*}[ht!]
    \centering
    \includegraphics[width=\linewidth]{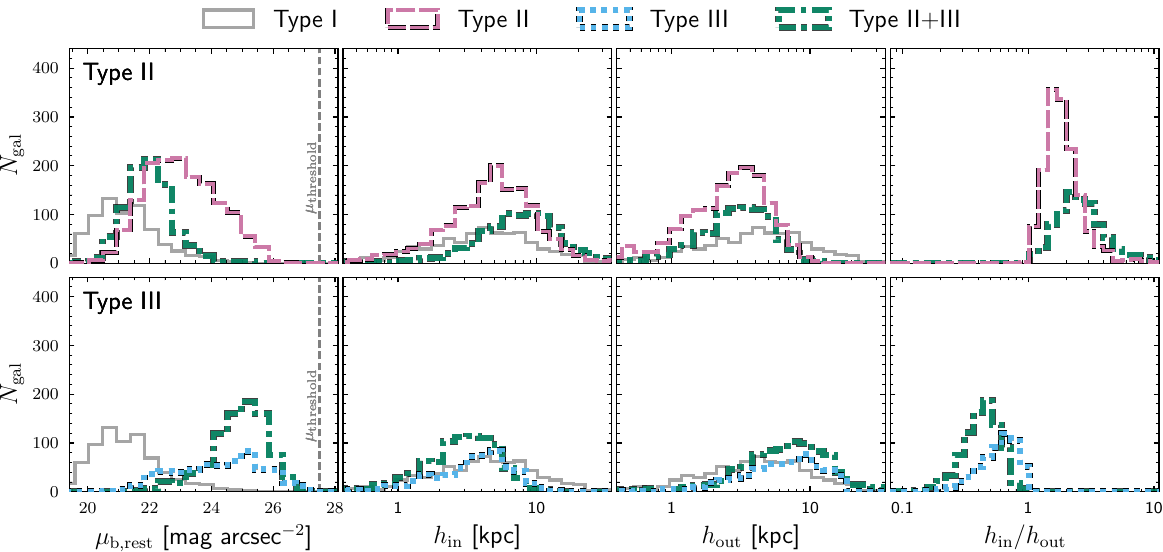}  
    \caption{Distributions of key structural parameters for galaxies with exponential disc breaks: surface brightness at the break radius (\textit{first} column), inner scale length (\textit{second} column), outer scale length (\textit{third} column), and the ratio of inner to outer scale lengths (\textit{fourth} column). The panels are arranged by break type: Type~II profiles are shown in the \textit{first} row, and Type~III profiles in the \textit{second} row. Different break classes are distinguished by colour and pattern: Type~II (purple dashed curves), Type~III (blue dotted curves), and Type~II+III (green dashed--dotted curves). The surface brightness level and scale~length of Type~I (in gray curves) profiles are also shown for reference. In the \textit{first} column, the value shown for Type~I corresponds to the central surface brightness of the exponential disc ($\mu_0$). For composite Type~II+III breaks, the individual Type~II and Type~III breaks are plotted separately in the corresponding rows.}
    \label{fig:EUC_breaks_histograms}
\end{figure*}

We investigate the distribution of the parameters for the most common disc breaks types, from our reliable sample of 4385 galaxies. 
Figure~\ref{fig:EUC_breaks_histograms} presents, from left to right, the distribution for the surface brightness at the break radius, the inner and outer scale~lengths, and the ratio between scale~lengths, often called break strength, for Type~II (first row) and Type III (second row) profiles. For profiles with multiple breaks, such as Type~II+III, we show the distribution of the Type~II and Type~III breaks separately. 

Disc breaks occur at a wide range of surface brightness levels, between 21 and 27 mag\,arcsec$^{-2}$ in $\mu_{I_{\rm E}}$. However, the median break surface brightness differs significantly between types: Type~II breaks occur at brighter levels ($\sim23.0\,$mag\,arcsec$^{-2}$) than Type~III breaks ($\sim24.2\,$mag\,arcsec$^{-2}$). The Type~II+III profiles have even brighter Type~II break surface brightnesses (around $\sim21.9$\,mag\,arcsec$^{-2}$), while the Type~III breaks are marginally fainter ($\sim24.9$\,mag\,arcsec$^{-2}$) than their counterparts.

The distributions of inner and outer scale~lengths reflect the expected broad distributions due to their correlation with stellar mass \citep{Laine14}.
While the inner scale lengths show relatively small variations across break types, the outer scale lengths exhibit systematic differences. Type~III discs tend to have larger outer scale lengths than Type~II discs, due to a similar behaviour observed in their respective components in Type~II+III profiles.
We also find that the inner scale lengths of Type~II components in Type~II+III galaxies are on average 15\% longer than in pure Type~II discs. Conversely, the inner scale lengths of Type~III components in Type~II+III are about 5\% shorter than those in pure Type~III discs. 
As a result, the break strength is typically enhanced in Type~II+III galaxies, implying more pronounced radial transitions than in the single-break analogues.

\subsection{\label{sec:EUC} Break location}
\begin{figure*}[ht!]
    \centering
    \includegraphics[width=1\linewidth]{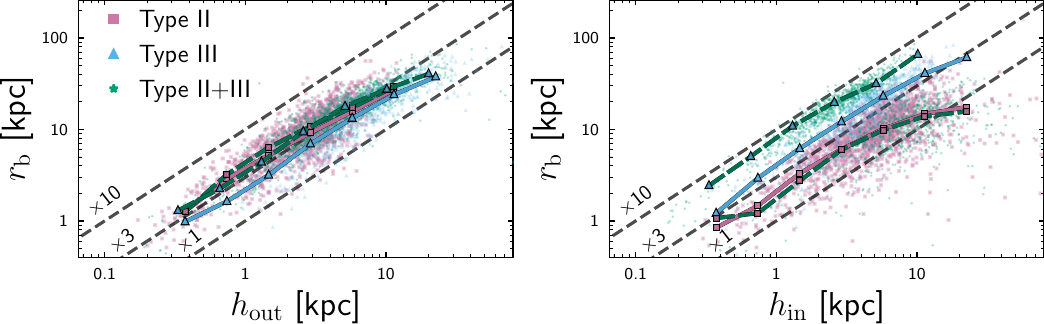}
    \caption{Break radius as a function of the inner (\textit{left} panel) and outer (\textit{right} panel) disc scale lengths. Each break type is colour-coded according to the scheme in Fig.~\ref{fig:EUC_types}. Both axes are shown on a logarithmic scale. Dashed lines indicate fixed ratios between the break radius and the corresponding scale length ($r_{\rm b}=n\, h_{\rm in,out}$ with $n=1,3,10$). Solid curves with markers show the mean trend for each break type. Individual breaks of Type~II+III profiles are shown independently using green dashed curves with purple squares and blue triangles for the Type~II and Type~III, respectively. }
    \label{fig:rbreak_vs_h}
\end{figure*}

We present the position of the radius with respect to the scale length of the disc in Fig.~\ref{fig:rbreak_vs_h}. We find that there is a tight correlation between the outer scale length and the position of the breaks, almost irrespective of the break type.  
The distribution of the outer ratio $r_{\rm b}/h_{\rm out}$ follows a log-normal distribution, and 75\% of breaks occur at $2.9^{+1.7}_{-1.1}$ outer scale~lengths. The upper and lower values are estimated as the difference between the percentiles 87.5 and 50, and 50 and 12.5, respectively.

In contrast, the break radius in units of the inner scale length shows different trends for Type~II and III galaxies. 
In terms of the inner scale~length, Type~II breaks occur at smaller scale~lengths than those in Type~III. Both distributions also follow log-normal distributions centred at $1.7^{+1.1}_{-0.9}$ and $5.6^{+1.7}_{-2.4}$. The uncertainty bars are set to include 75\% of the breaks. Subtle differences are also seen between types in relation to the outer scale~length, where Type~III breaks occur more inside than Type~II breaks when compared to the outer scale~length.

\subsection{\label{sec:EUC_results_break_evolution} Break type evolution}

Thanks to the large sample size and broad redshift coverage, we can assess the evolution of disc break types across time. Figure~\ref{fig:EUC_type_vs_z} shows the relative fraction of each break type as a function of lookback time, binned in logarithmic steps in redshift. Within each bin, we compute the fraction of each profile type, correcting for classification uncertainties based on the estimated $\sim70\%$ accuracy of our automated method.

To quantify the systematic uncertainties and propagate classification errors, we perform 1000 iterations of a bootstrap resampling with replacement. In each iteration, we randomly replaced 30\% of the galaxy classifications (1120 galaxies) based on the observed distribution of break types in the local Universe. We model this prior distribution as a normal distribution with mean and standard deviation values of $(13\pm7)\%$, $(52\pm6)\%$, $(25\pm4)\%$, and $(10\pm4)\%$ for Type~I, II, III, and II+III profiles, respectively. This selection is done by combining the results from \cite{Laine14}, \cite{Pranger17}, and \cite{Xu24}. Additionally, to account for redshift uncertainties, we perturb the redshift values of each galaxy with a Gaussian scatter of 20\%. For each resampling, we measure the fraction of types at each redshift bin. We also investigate the effects of the extended point spread function (ePSF) of \Euclid in high-redshift galaxies in Appendix~\ref{app:EUC_PSF}. We find that the extremely good ePSF of \Euclid does not affect the classification of the profiles for our sample of galaxies. 

The data shown in Fig.~\ref{fig:EUC_type_vs_z} correspond to the mean values over all bootstrap realisations. The error bars reflect the quadratic sum of the standard deviation from the bootstrap samples and the binomial error for each redshift bin, computed as $\sqrt{N_{\rm gal}(\mathrm{type},z)}/N_{\rm gal}(z)$, where $N_{\rm gal}(\mathrm{type},z)$ is the number of galaxies for each type within each redshift bin, and $N_{\rm gal}(z)$ is the number of galaxies within each redshift bin. 

We observe a clear evolutionary trend in the fractions of the break types over time. Type~II profiles, which dominate the local Universe with fractions exceeding $50\%$, are progressively less frequent at higher redshifts. Beyond a lookback time of $\sim3$\,Gyr ($z\gtrsim0.5$), Type~III and Type~II+III profiles surpass Type~II in frequency. Both Type~III and II+III profiles exhibit a steady increase with redshift, rising from $\sim10\%$ at $z\sim0.05$ to nearly $30\%$ at $z\sim1$. 
We also observe a mild decreasing trend in the fraction of Type~I profiles with increasing redshift. However, this variation is considerably weaker than that observed in other types and may be consistent with a roughly constant behaviour over time. 

We also include in Fig.~\ref{fig:EUC_type_vs_z} the values from the literature as a reference for the results obtained in previous studies. We recover values similar to those in the previous studies for low redshift. The fraction of Type~II profiles are in agreement with most of the studies, while the fractions of Type~I and III profiles are more similar to those in isolated regions \citepalias{Sanchez-Alarcon23} rather than to the field or to the cluster environments \citep{Pranger17}. For our intermediate redshift ($z\sim0.5$), we found excellent agreement between values reported by \citet{Borlaff17} for Type~III and Type~II, while we see large difference for Type~I. Lastly, for larger redshift ($z\sim1$), we see some discrepancies for Type~I and II profiles fractions with respect to those reported by \cite{Xu24}.

\section{\label{sec:EUC_discussion}Discussion}
\subsection{Properties of disc breaks}

Figure~\ref{fig:EUC_breaks_histograms} shows the scale~lengths of the discs and surface brightness levels at the break radii for the various surface brightness profile types. Galaxies show a wide variety of inner scale lengths without any significant difference between disc types. The main driver of this large variety is the galaxy stellar mass, as the mass and size of galaxies are closely linked \citep[e.g.][\citetalias{Sanchez-Alarcon25}]{Laine14, Munoz-Mateos15, Trujillo20}. 
%The structural analysis also confirms that the outer scale lengths of Type~III profiles are systematically longer than those of Type~II discs, which was reported by many studies in the literature \citep[e.g.][]{Laine14,Pranger17}, adding confidence to our analysis. 
This points to different formation channels for the two types of breaks: Type~II (down-bending) profiles likely reflect internal processes introducing breaks at brighter levels such as radial migration and angular momentum redistribution within a stable disc \citep{Debattista06, Roskar08}, while Type~III (up-bending) profiles are more sensitive to environmental effects, accretion events, and possibly reformed star-forming outer discs \citepalias{Watkins19,Sanchez-Alarcon23}. The latter affects mostly the outskirts of galaxies, introducing breaks at fainter levels.

We find that break radii strongly correlate with outer disc scale lengths, typically occurring at $r_\mathrm{b} \sim 3\,\,h_\mathrm{out}$ for all break types. This result aligns with previous studies in the local Universe  \citep[e.g.][]{Pohlen06,Laine14} and simulations \citep{Minchev12}, but the \Euclid sample extends this conclusion over a broader redshift range and with a far greater statistical power.

The surface brightness level at the break radius varies with break type. Type~II profiles occur at levels around $1.2\,\mathrm{mag\,arcsec}^{-2}$ brighter than Type~III profiles, consistent with their respective associations to outer star formation thresholds and extended low surface brightness components. Interestingly, galaxies with combined Type~II+III profiles exhibit enhanced break strengths and more extreme surface brightness transitions. This suggests a superposition of both internal and external processes, such as secular evolution shaping the inner disc and minor mergers or halo growth influencing the outskirts.

We explore the possible relations with stellar mass and redshift (Appendix~\ref{app:EUC_Mass}), but find only mild trends, similar to those encountered for the whole sample, which could be biased by the mass and redshift relation of our sample. We are thus prevented from drawing any firm conclusions. We leave this analysis for a further data release of \Euclid, when we will have a larger sample available. Despite this limitation, this work represents the most detailed study of disc breaks to date. 

Overall, the structural properties of disc breaks support a picture in which the outer disc carries the imprint of both environmental history and secular growth, while the inner disc retains a more uniform scaling. The ability of \Euclid to measure these properties down to surface brightness levels of $\sim29.5\,\mathrm{mag\,arcsec}^{-2}$ at $z\sim1$ is key to disentangling these overlapping evolutionary processes.

\subsection{Evolution of the frequency of disc break types}

The evolution in the frequency of break types provides insights into a shift in the dominant physical mechanisms shaping disc galaxies across time. In the low-redshift Universe ($z \lesssim 0.2$), down-bending Type~II profiles are the most common, consistent with a scenario where discs are well-settled, gas fractions are lower, and secular processes such as bar-driven migration dominate \citep{Q1-SP069}. However, at earlier epochs ($z \gtrsim 0.5$), the significantly increased prevalence of Type~III and Type~II+III profiles suggests a transition in the dominant disc-building processes.

\begin{figure*}[ht!]
        \centering
    \includegraphics[width=1\linewidth]{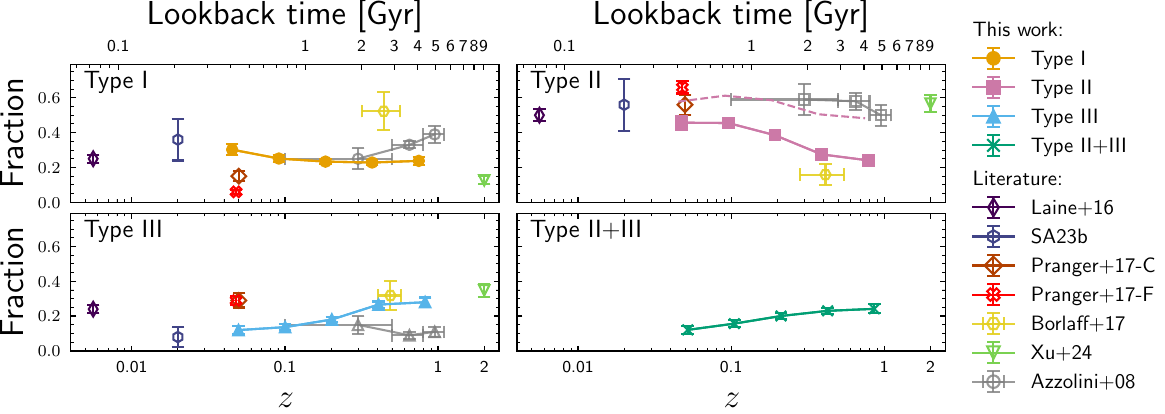}
    \caption{Evolution of the frequency of disc break types with time. The fraction of Type~I (\textit{top left}), Type~II (\textit{top right}), Type~III (\textit{bottom} \textit{left}), and Type~II+III (\textit{bottom} \textit{left}) galaxies is shown with respect to the redshift (\textit{bottom} \textit{x}-axis) and lookback time (upper \textit{x}-axis). The relation from this work is shown with curves, lines, and filled symbols. Literature values from \cite{Azzollini08}, \cite{Laine14}, \citetalias{Sanchez-Alarcon23}, \cite{Pranger17}, \cite{Borlaff17}, and \cite{Xu24} are shown with empty symbols, and colour coded according to the legend on the right.}
    \label{fig:EUC_type_vs_z}
\end{figure*}

One could argue that the observed trend can be influenced by the mass dependency on the redshift in our sample. However, studies in the literature, such as \cite{Laine16}, indicate that Type~II and Type~III fractions increase with stellar mass by approximately 15\% for stellar masses ranging from $8.7$ to 11 in $\log_{10}(M_*/M_\odot)$. Based on this study, we would expect to see a similar increase of around 15\% in Type~II and Type~III fractions with redshift. In contrast, we observe the opposite trend in Type~II profiles. In Appendix~\ref{app:EUC_Mass}, we explore whether the evolution of Type~III fraction can be attributed to the mass bias in the sample. We find that the trends are most likely driven by redshift for all types except for Type~II+III, where the trend is compatible with being driven by the stellar mass. Therefore, we conclude that these trends are indicative of a genuine physical evolution occurring in galaxies. 

The evolution of the break fraction with time can be interpreted in terms of the cosmic history of galaxy environment and star formation. At higher redshifts, galaxies are more gas-rich, dynamically disturbed, and are embedded in denser environments than nowadays \citep{Duncan19}. These conditions favour the formation of extended, low surface brightness components, either through outer disc star formation fuelled by gas accretion, or through structural heating via interactions and mergers. 

Conversely, the increase in the Type~II profile fraction with decreasing redshift suggests that the processes producing down-bending discs, such as star formation thresholds linked to gas surface density or angular momentum cutoff, are more efficient at later times, when galaxies are dynamically more relaxed and their gas reservoirs are increasingly depleted. This matches the expectation that secular evolution becomes more dominant as time progresses \citep{Kormendy04}. This trend compares well with the results described in other Euclid Q1 papers, such as \cite{Q1-SP043} where they confirmed a decrease in the frequency of barred galaxies as lookback time increases, which can play a role in the evolution of the Type~II profiles, and in \cite{Q1-SP069}, where they found that the morphological transformation of field galaxies is mainly driven by secular processes taking place while the galaxies are still actively star-forming on the main sequence, supporting the scenario of increasing Type~II with time.

The mild evolution of Type~I profile fractions can be interpreted through two different scenarios. It either suggests that a subset of discs remain structurally unperturbed over long timescales, or that Type~II profiles are transformed to Type~I by accreting material \citep[similar trends are found in galaxy clusters by][]{Mondelin25}. The former scenario may be associated with isolated environments or galaxies with specific angular momentum and gas accretion histories that avoid the instabilities leading to breaks. The latter scenario can be associated with an increase in their merger rate at higher redshifts, which will explain the subtle curvature transition, moving from a negative slope to a slight positive gradient in Type~I fraction at higher redshift \citep[also seen by][]{Azzollini08}. At redshifts $z>0.5$, the environment is denser, and galaxies are more likely to enter higher-density environments, where they could lose their Type~II breaks. 

Another explanation for the subtle curvature transition of the fraction of Type~I profiles and the decrease of that of Type~III (at $z<0.2$) could be the transformation of Type~III profiles into Type~I ones. This transformation may occur as the outer excess light characteristic of Type~III discs fades due to the ageing of stellar populations and the quenching of star formation in the outer disc. The observed colour profiles that show redder and older stellar populations in the outer regions of many Type~III discs support this scenario \citep{Bakos08,Watkins19}.

In comparison with the literature, we observe some discrepancies between our fraction of breaks and the values reported by \cite{Yu25} and \cite{Azzollini08} for high-redshift galaxies. We find a lower frequency of Type~II profiles and a different fraction of Type~I profiles. 
Neither \cite{Yu25} nor \cite{Azzollini08} model multiple discs, so we combine Type~II+III profiles into a single Type~II category for a fair comparison (shown with the dashed pink curve in Fig.~\ref{fig:EUC_type_vs_z}). The decrease in Type~II profiles remains evident, with a less pronounced trend, but we find values compatible with those of the \cite{Yu25} and in good agreement with the trend found by \cite{Azzollini08}, indicating that approximately 50\% of profiles are of Type~II at $z\sim1$. The limited size of \cite{Yu25} and \cite{Azzollini08} samples (247 and 435 galaxies, respectively) may help cause the differences found in our results for the other types of breaks.

For Type~III profiles, we find excellent agreement with the values reported by \citet{Borlaff17} and \citet{Xu24}, but a significant discrepancy with those from \citet{Azzollini08} who used HST data from the GOODS-S field, reaching surface brightness limits of $\sim 28 \,$mag~arcsec$^{-2}$. Because these data are shallower and potentially suffer from sky over-subtraction effects \citep[][]{Borlaff19}, \citet{Azzollini08} likely missed most of the Type~III breaks, which typically appear at fainter surface brightness levels (Fig.~\ref{fig:EUC_breaks_histograms}). This observational bias may also explain the discrepancies observed in the fraction of Type~I profiles, as undetected Type~III breaks would naturally be misclassified as Type~I profiles. Furthermore, while \citet{Borlaff17} also utilized HST data (from the GOODS-N field), their study focused exclusively on lenticular galaxies which could further drive the differences we observe in the Type~I profiles.

A significant source of uncertainty in the measured break type frequencies is the underlying environment probed at each redshift. Recent studies suggest that sparser environments tend to favour Type~I profiles while suppressing the frequency of Type~III discs \citepalias{Watkins19, Sanchez-Alarcon23}. The relatively high frequency of Type~I profiles in our low-redshift sample might be due to us sampling preferentially more isolated environments, consistent with the trends observed in the isolated galaxy sample of \citetalias{Sanchez-Alarcon23}. Despite this potential bias, the evolution of Type~II frequencies appears robust, as their frequency is less sensitive to environmental density. Moreover, the large number of galaxies in each of our redshift bins (exceeding 200 in all cases) mitigates environmental differences with the average number of galaxies at each epoch, which exceeds the sample sizes of previous studies. In future work, using the forthcoming \Euclid data releases will allow us to trace the representative densities at each epoch to a much higher degree of precision by increasing the sample size and area coverage.

These preliminary results come with a caveat, which is the mass biases at lower redshift (missing massive galaxies) due to the limited area coverage. This introduces uncertainties within the evolutionary trend presented. The increase in stellar mass with redshift would enhance the positive evolutionary trend of Type~III and Type~II+III profiles \citep[as seen in][]{Laine14}. However, this trend persists even when we limit the stellar mass range (Appendix~\ref{app:EUC_Mass}). Furthermore, the similarities observed in the fractions in our low redshift sample and those from \citetalias{Sanchez-Alarcon23} suggest that the limited area coverage in Q1 might be tracing more isolated environments at low redshift. This bias could also enhance the positive trend of Type~III breaks with redshift, as isolated environments suppress Type~III breaks \citepalias{Sanchez-Alarcon23}. 

Taken together, these trends reflect the interplay between environmental factors and internal evolution. The dominance of Type~III profiles at high redshift highlights the importance of external processes in early disc assembly, while the emergence of Type~II structures at later times underscores the increasing role of internal, secular disc evolution. Our results favour a scenario where, as time evolves, galaxies on average experience fewer environmental effects and discs can evolve to become more stable, Type~II+III profiles can then transform into Type~II profiles, which offers a coherent structural counterpart to the well-known decline in cosmic star formation rate density and morphological transformation of galaxies over the past 10\,Gyr.

%Current efforts in the \textit{Euclid} consortium are focused on characterising the environment and the structure of the cosmic web \citep[][]{Q1-SP028}. These measurements will be incorporated to study the traced environments at each epoch. However, these biases do not significantly affect the evolutionary trends in Type~II profiles, which are less influenced by the environment. Our analysis in Appendix~\ref{app:EUC_Mass} suggests that we can have greater confidence in the results regarding the evolution of Type~II, Type~III and Type~I profiles than those of Type~II+III ones.

%Moreover, we are currently testing SED algorithms to infer physical properties, such as stellar masses, which will help reduce uncertainties in the stellar masses of the galaxies. These biases will be analysed in further detail prior to the publication of the results.

%Future data releases will include large areas which will also reduce this bias. A larger statisticall sample will allow to study the evolutionary trend at each environment. 

%Further analysis of subsequent data releases from \Euclid can strengthen these findings by increasing the number of galaxies and the survey area. The new data will enable investigation of mass degeneracy and environmental influences at each epoch.

\section{\label{sec:EUC_conclusions}Conclusions}

We develop an automatic surface brightness profile analysis pipeline to produce reliable masks, deep surface brightness profiles, structural and photometric parameters, and surface brightness profile classifications. Our pipeline is optimised for analysis at low surface brightness levels. We perform a detailed analysis of the structural properties and the evolution of disc breaks in a sample of 4385 disc galaxies up to $z=1$ using images from the Euclid Quick Release 1 (Q1). This work represents the most comprehensive statistical study of disc breaks to date over such a wide redshift range. Our analysis provides new constraints on the physical processes shaping disc structure over a large fraction of the age of the Universe. 

The main conclusions of this work are the following:

\begin{itemize}
    \item We classify disc galaxies into up to four disc components based on automated piecewise modelling of their surface brightness profiles. Visual inspection of a subsample highlights that our automatic method yields an accuracy of 70\%. Among a reliable sample of 4385 galaxies, we find the following profile type fractions: Type~II (31\%), Type~II+III (21\%), Type~I (16\%), and Type~III (16\%).

    \item The break radius $r_\mathrm{b}$ scales tightly with the outer disc scale length $h_\mathrm{out}$, with 75\% of breaks happening at ratios $r_\mathrm{b}/h_\mathrm{out} = 2.9^{+1.7}_{-1.1}$, nearly independent of break type. This suggests a universal scaling relation in the exponential outer discs.

    \item Surface brightness levels at the break radius differ systematically across types: Type~II breaks occur at brighter levels ($\sim23.0$ mag\,arcsec$^{-2}$) than Type~III ($\sim24.2$ mag\,arcsec$^{-2}$). Combined Type~II+III profiles show an enhanced break strength.

    \item Inner scale lengths are relatively consistent across types, while outer scale lengths are significantly larger for Type~III and Type~II+III profiles, consistent with the latter occurring in extended or environmentally influenced outer discs.

    \item We detect significant evolutionary trends in the frequency of break types with redshift. Type~II profiles, dominant at low redshift, exhibit lower prevalence at $z > 0.5$, while Type~III profile fractions increase steadily with higher redshifts, each reaching $\sim30\%$ at $z \sim 1$. This evolution reflects a transition from environmentally driven disc growth at higher redshifts to secular processes at lower redshifts.

    \item Type~I profiles show only mild redshift evolution, suggesting that a fraction of discs either remains structurally undisturbed over time, or alternatively undergoes a steady transformation from Type~II and III profiles to Type~I, introducing subtle fluctuations in the fraction of these profiles.

    %\item The large number of galaxies in each redshift bin ($N > 200$) ensures statistical robustness. Bootstrap resampling confirms that classification uncertainties do not drive the observed trends.

    \item The similarity of our low-redshift Type~III fractions to those found in isolated environments suggests that our sample may over-represent sparse environments at low redshift. Additionally, the relationship between mass and Type~III profiles could lead to a positive trend due to the dependence of mass on redshift in our sample. These two observational biases may be influencing the evolution of Type~III profiles. However, the trends for Type~II profiles appear to be robust against this bias.

\end{itemize}

Our automated pipeline successfully models 70\% of the sample, delivering surface brightness profiles, photometric parameters, and reliable break classifications for the entire sample. The pipeline is modular, scalable for upcoming full-depth \Euclid data releases, and will be public.
Future \Euclid data will enable even deeper and more representative studies across cosmic volume, allowing further exploration of the role of environment, mass, and morphology in disc break formation and evolution.

\section{Data and code availability}
The catalogues produced with the methods described in this work will be available at the CDS via anonymous ftp to \url{cdsarc.cds.unistra.fr} or via \url{https://cdsarc.cds.unistra.fr/viz-bin/cat/J/A+A}. Additionally, the pipeline is fully available in this repository \url{https://github.com/PabloMSanAla/euclid-surface-photometry} and is completely implemented and functional in the Datalabs\footnote{\url{https://datalabs.esa.int/}} platform from ESA.

\begin{acknowledgements}
%\AckERO  
\AckEC  
\AckQone

We acknowledge support from the Agencia Estatal de Investigaci\'on del Ministerio de Ciencia, Innovaci\'on y Universidades (MCIU/AEI) under the grants “The structure and evolution of galaxies and their outer regions” and the European Regional Development Fund (ERDF) with references PID2019-105602GBI00/10.13039/501100011033 and PID2022-136505NB-I00/10.13039/501100011033. 
Co-funded by the European Union (MSCA Doctoral Network EDUCADO, GA 101119830 and Widening Participation, ExGal-Twin, GA 101158446). 
PMSA wishes to acknowledge the contribution of the IAC High-Performance Computing support team and hardware facilities to the results of this research.  
PMSA acknowledges that part of this research was sponsored by the National Aeronautics and Space Administration (NASA) through a contract with ORAU. The views and conclusions contained in this document are those of the authors and should not be interpreted as representing the official policies, either expressed or implied, of NASA or the U.S. Government. The U.S. Government is authorized to reproduce and distribute reprints for Government purposes notwithstanding any copyright notation herein.
JR acknowledges financial support from the Spanish Ministry of Science and Innovation through the project PID2022-138896NB-C55.
SC acknowledges funding from the State Research Agency (AEI) of the Spanish Ministry of Science, Innovation, and Universities under the grant “The relic galaxy NGC 1277 as a key to understanding massive galaxies at cosmic noon” with reference PID2023-149139NB-I00. 
\end{acknowledgements}

%
% Here comes the reference list, generated via bibtex from
% your bibfile my.bib and Euclid.bib. Please make sure that
% the same paper is not referenced twice, one from your my.bib
% file, and once from Euclid.bib.
%

\bibliography{references} % add my.bib, containing your bibentry file 

%
% Now you can add appendices.
% In this example, the appendices are in one column mode.
% If that is not requires, comment out \onecolumn
% Note that appendices in A\&A come {\it after\/} the references.

\begin{appendix}

\section{\label{app:EUC_MER}Comparison with MER}

To validate the photometry of the pipeline, we compare our integrated magnitudes with those derived using MER \citep{Q1-TP004}. MER provides a wide variety of ways to measure the photometry of a source. In addition to fixed aperture photometry, MER also provides total flux estimates (\texttt{FLUX\_DETECTION\_TOTAL} column) based on the Kron radius of the detected sources on the \IE image. Then, to extract the total flux in the other bands (\YE, \JE, \HE, and \texttt{EXT}), the photometry cookbook\footnote{\url{http://st-dm.pages.euclid-sgs.uk/data-product-doc/dm10/merdpd/merphotometrycookbook.html}} of \Euclid suggests scaling the total flux using a colour correction using a colour correction from aperture photometry (hereafter \texttt{APHOT}) or via \texttt{TEMPLFIT} \citep[][; hereafter \texttt{TPHOT}]{Merlin15}. The \texttt{APHOT} colour correction relies on the aperture defined in terms of the FWHM. In contrast, the \texttt{TPHOT} colour correction uses priors from higher resolution images (VIS) to measure the photometry in other bands. The scaling factor for the \HE band is
\begin{equation}\label{eq:mer_aperture}
    \frac{\rm \texttt{FLUX\_H\_4FWHM\_APER}}{\rm \texttt{FLUX\_VIS\_4FWHM\_APER}} \;,
\end{equation}
for the \texttt{APHOT} correction using the largest aperture ($4\,$\,FWHM), while for \texttt{TPHOT} it is 
\begin{equation}\label{eq:mer_templfit}
    \frac{\rm \texttt{FLUX\_H\_TEMPLFIT}}{\rm \texttt{FLUX\_VIS\_TO\_H\_TEMPLFIT}} \;.
\end{equation}
These factors are multiplied by the parameter \texttt{FLUX\_DETECTION\_TOTAL} to obtain the total flux of the source in the NIR and \texttt{EXT} bands\footnote{\texttt{EXT} acronym refers to the complementary optical images included in EWS from external ground-based facilities.}. 
These are the best estimates of the total flux of the source in MER. 

\begin{figure*}[ht!]
    \centering
    \includegraphics[width=\linewidth]{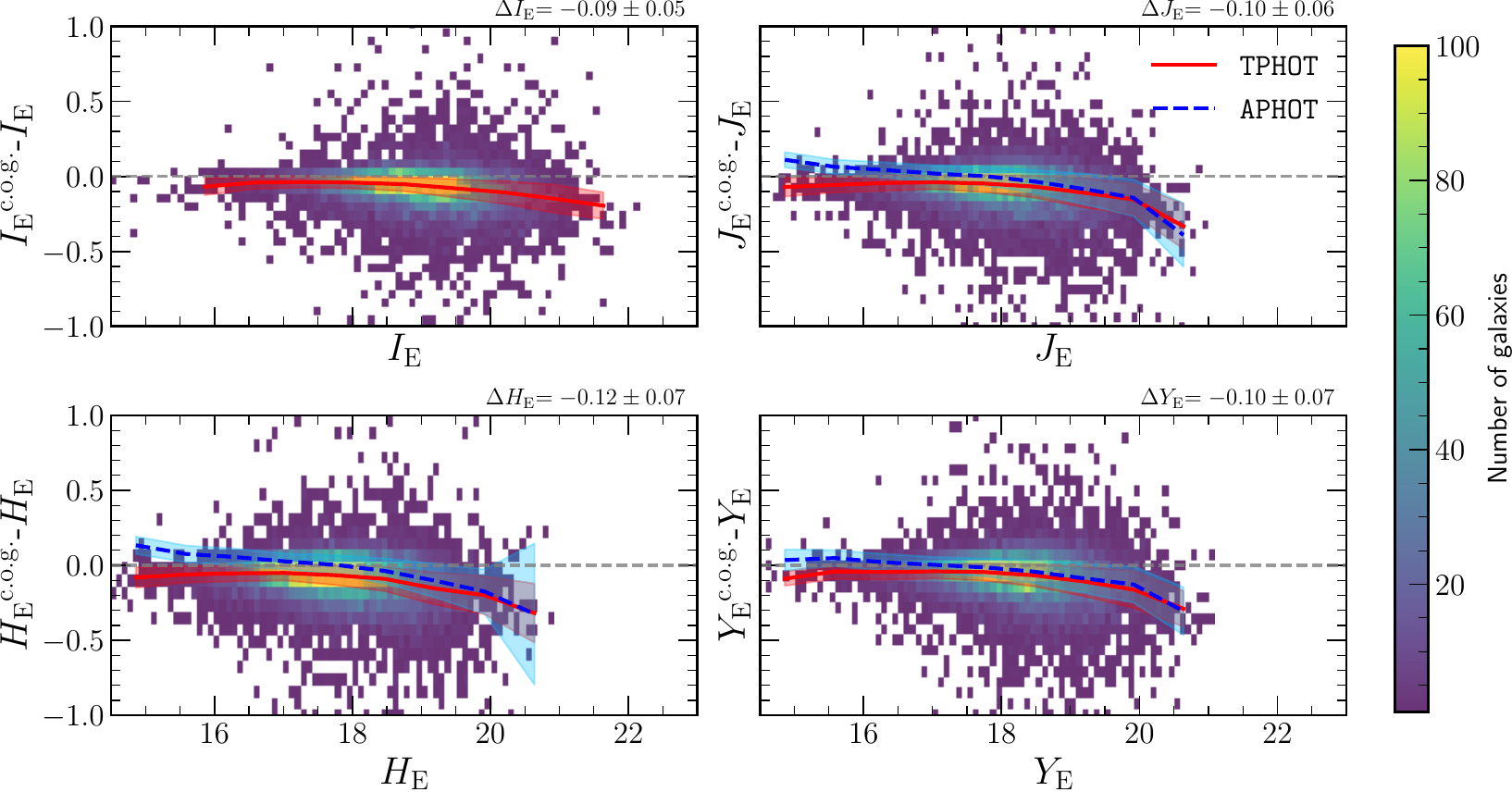}
    \caption{Comparison of integrated magnitudes measured with MER using the \texttt{TEMPLFIT} colour correction. All panels show the difference in magnitudes for our sample of galaxies (shown with a density plot) with respect to the \texttt{TPHOT} magnitude. Each panel shows a different \Euclid band. The c.o.g. label refers to our asymptotic magnitudes measured using the curve of growth (Sect.~\ref{sec:EUC_M4}). The grey line indicates a null difference. The red curve shows the average of all values within bins of 0.6 magnitudes, and the grey filled regions show the mean root mean square RMS difference of all values and the average curve. The average difference and RMS are shown at the \textit{top right} of each panel. The blue dashed curve shows the average trend for \texttt{APHOT} colour correction.  } 
    \label{fig:EUC_MER_comparison}
\end{figure*}

We show the comparison between our asymptotic magnitudes (Sect.~\ref{sec:EUC_M4}) and the \texttt{TPHOT} apparent magnitudes for \Euclid's bands, \IE, \YE, \JE, and \HE in Fig.~\ref{fig:EUC_MER_comparison}. Each panel shows the difference between our determination of the magnitudes and that of the \texttt{TPHOT} colour correction. The median trend is shown with the red curve. The blue curves show the median trend for the difference between our magnitudes and the \texttt{APHOT} photometry with MER. 

We find good agreement between the values reported by MER and ours, with some differences that depend on the selected colour correction.  In the case of \texttt{TPHOT}, we systematically recover brighter magnitudes than MER, with a mean magnitude difference of $-0.09\pm0.05$, $-0.10\pm0.07$, $-0.10\pm0.06$, $-0.12\pm0.07$, and  for \IE, \YE, \JE, and \HE respectively. The error is estimated as the root mean square (RMS) of the difference between all values and the mean trend curve. This difference tends to increase with the magnitude of the source, reaching $\sim0.2\,$mag at levels of $21\,$mag in \IE and at $19\,$mag for the NIR bands. 
In the case of \texttt{APHOT} photometry (blue curves) of NISP bands (\YE,  \JE, and \HE), we find similar trends for the three bands with some correlation with the magnitude of the sources. For sources brighter than $17.5\,$mag, our asymptotic magnitudes tend to be fainter than the \texttt{APHOT} magnitudes (around $\sim0.05\,$mag). For sources fainter than $17.5\,$mag, we recover brighter magnitudes than \texttt{APHOT} ($\sim 0.1\,$mag).  

These differences can be associated with the systematics in the different methodologies used to measure the integrated magnitudes and with the background subtraction calibration. Our method consists in integrating the surface brightness profiles that extend out to the region where the background is reached and extrapolating the asymptote. This technique offers a correction on the flux missed by the lack of depth of the data, and we also account for possible masked regions within the galaxy. 
In addition, we include the MER background maps and recompute the background using our own methods (Sect.~\ref{sec:EUC_M3_bkg}), which may also introduce flux of the galaxy previously removed by oversubtraction effects.
This flux correction and the addition of the background maps could explain the brighter sources with respect to \texttt{TPHOT} photometry. However, with respect to \texttt{APHOT} photometry, NISP shows two different behaviours. For the faintest sources, our magnitudes recover more flux, as expected from the same reasoning as with \texttt{TPHOT}. For brighter sources, \texttt{APHOT} photometry recovers brighter sources than our method. The magnitudes for the NISP instrument are scaled using the total flux detected in \IE and a normalisation factor (Eqs.~\ref{eq:mer_aperture} and \ref{eq:mer_templfit}) from the fraction between the flux measured in the \IE and the NIR bands that depend on the method. If the source is brighter, concentrated, and large enough, the aperture only covers the central region of the source, and the correction extrapolates the colour from the central region to the outskirts of the source. In most cases of galaxies, there are gradients in colours, and the normalisation term cannot reproduce the colour of the outskirts. The effect is more prominent on the brightest and largest galaxies. 

\begin{figure*}[ht!]
\centering
    \includegraphics[width=0.9\linewidth]{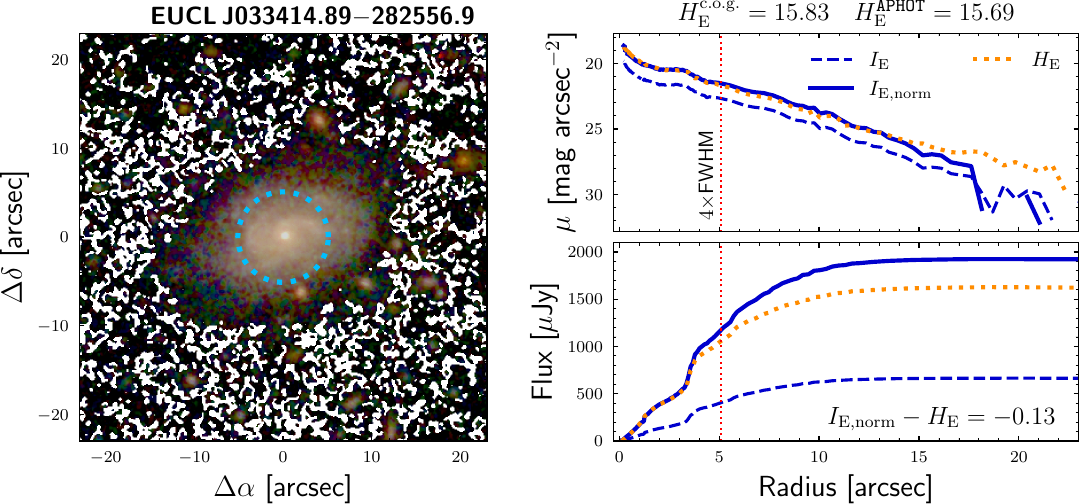}
    \caption{Example of the \texttt{APHOT} colour correction on a galaxy of our sample and the surface and cumulative brightness profile. The \textit{left} panel shows an LRGB image with the aperture of $4\,$FWHM shown with the blue circle. The \textit{top right} panel shows the surface brightness profile measured for \IE (blue dashed curve), \HE (orange curve), and the normalised \IE bands using the aperture colour correction from MER. The \textit{bottom} panel shows the curve of growth for those profiles. The vertical red dotted line shows the radius of the aperture. The asymptotic magnitudes and the aperture-corrected magnitudes from MER are shown above the \textit{top right} panel. The difference between the normalised \IE and the \HE magnitudes is shown in the \textit{bottom} \textit{right} panel.   }
    \label{fig:EUC_MER_profiles}
\end{figure*}

We show an example of a galaxy with a brighter magnitude given by \texttt{APHOT} in Fig.~\ref{fig:EUC_MER_profiles}. The left panel shows an LRGB image and the aperture of size $4\,$FWHM used to measure the \texttt{APHOT} colour correction (Eq.~\ref{eq:mer_aperture}). The top right panel shows the surface brightness profile for \IE (blue dashed curve), \HE (orange curve), and the normalised \IE (blue curve) using the colour correction from MER. The bottom right panel shows the curve of growth for each profile. Around the location of the aperture, the profiles diverge, and the normalised profile overestimates the emission on the \HE band ($0.13\,$mag larger) due to a redder central region than the outskirts of the galaxy. This overestimation is due to tracing the colour of the central region of galaxies rather than the global colour, which explains the behaviour for bright sources ($> 17.5$ mag) for the \texttt{APHOT} colour correction. 

We measure the difference between \HE and the normalised \JE surface brightness profile for all galaxies and we find an average difference of ($0.07\pm0.06\,$) mag\,arcsec$^{-2}$ that reaches up to $0.12\,$mag\,arcsec$^{-2}$ for galaxies brighter than $17.5\,$mag. We advise to use the \texttt{TPHOT} colour correction, especially for bright and large sources. Additionally, our method recovers photometry with a magnitude scatter ($\sim0.05$--$0.07$) within the errors of the photometry provided by MER \citep[see ][]{Q1-TP004} but recovers slightly more flux ($\sim0.09$--$0.12$ magnitudes) than MER, especially for the fainter sources. 

\section{\label{app:EUC_PSF} Effect of the PSF on the profiles}
To account for the effect of the PSF on the surface brightness profiles, we investigate the difference between profiles affected by the PSF and free of PSF effects using galaxy models that are representative of the sample. We created galaxy models with a bright nucleus and an exponential profile using \texttt{IMFIT} \citep{IMFIT}. The models have profiles that resemble the profiles studied in Sect.~\ref {sec:EUC_results_break_evolution}. We create four model galaxies with Type~I, II, III, and II+III profiles. The parameters of the discs and the inclination used for the models were extracted from the smallest 10\% of galaxies (with $R_{25.5}<3.8\,$arcsec), as we expect them to be the most affected by the PSF \citep{Borlaff17}. 

\begin{figure*}[t!]
    \centering
    \includegraphics[width=\linewidth]{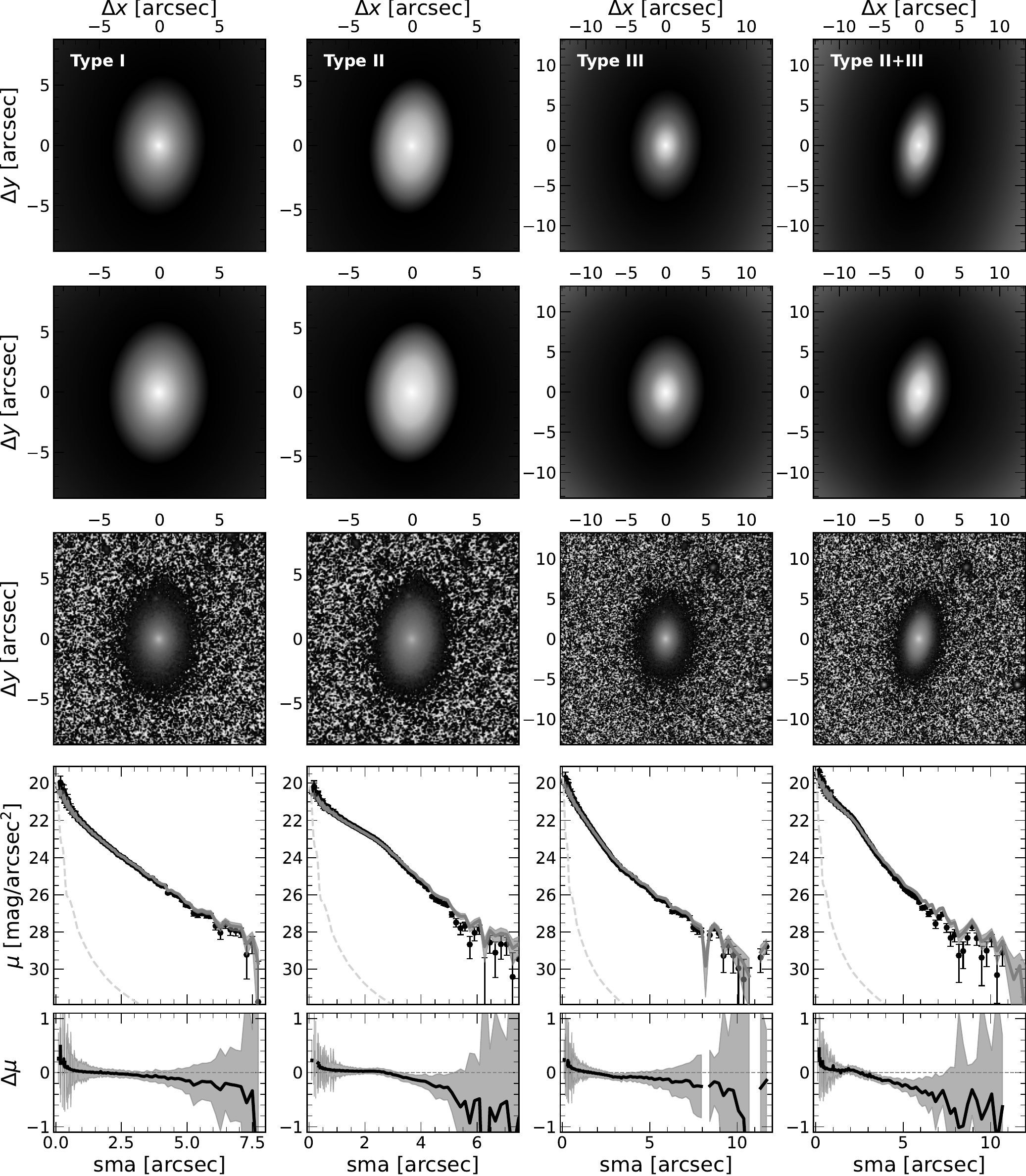}
    \caption{\small Effect of the PSF on the surface brightness profiles of simulated galaxies.  From left to right, each column represents a simulated galaxy for the most commonly studied disc profile types: Type~I, Type~II, Type~III, and Type~II+III.  
    From \textit{top} to \textit{bottom}, the panels include: the original model, the model convolved with the PSF, the model inserted in an \Euclid image, the surface brightness profiles, and the residuals.  
    In the surface brightness profiles panels, the black data points represent the model without PSF effects, while the grey line shows the convolved model. The grey dashed profiles indicate the PSF of \Euclid, normalised to the brightest point in the surface brightness profile of the simulated galaxy.}
    \label{fig:EUC_PSF}
\end{figure*}

Figure~\ref{fig:EUC_PSF} shows the models of the galaxies (first row), the models convolved with the PSF (second row), the convolved model injected in a \Euclid image on the \JE band (third row), the surface brightness profile (fourth row) of the models before (black points) and after (grey curve) the PSF convolution, and the residuals of the difference of both profiles (fifth row). We use the extended PSF measured in the ERO observations \citep{EROData}, and the surface brightness profile is shown with a dashed curve, scaled to the brightest pixel. 

The residual plots reveal some expected differences in the innermost regions, where the PSF has the strongest effect by redistributing central light. However, this redistribution does not significantly impact the extracted surface brightness profiles. At faint surface brightness levels ($\mu > 26$\,mag\,arcsec$^{-2}$), minor deviations become apparent in the residuals, but they remain lower than the error bars. The most affected case is the Type~II+III profile, likely due to the larger inclination of the selected galaxy model, which enhances PSF scattering in the outer regions. Despite these minor discrepancies, no significant structural changes are observed in the profiles that would alter their break classification.

To quantify the influence of the PSF on disc break characterisation, we re-analysed the profiles using M5 with a PSF convolution included. While the break classifications remain unchanged, the outer scale lengths are systematically affected, with values up to 20\% larger compared to those derived without PSF correction. 

We conclude that for galaxies of size larger than $R_{25.5}>3\,$arcsec the effect of the PSF on the profiles is negligible and does not affect the disc break classification. This will be the case in our study, as all the galaxies in our sample have an effective radius that is at least five times larger than the FWHM of the PSF of \Euclid. This ensures that the observed trends in Sect.~\ref{sec:EUC_results_break_evolution} are not an observational effect due to artificial breaks formed by the PSF.

\section{\label{app:EUC_Mass} Mass bias with redshift}
\begin{figure*}[ht!]
    \centering
    \includegraphics[width=0.9\linewidth]{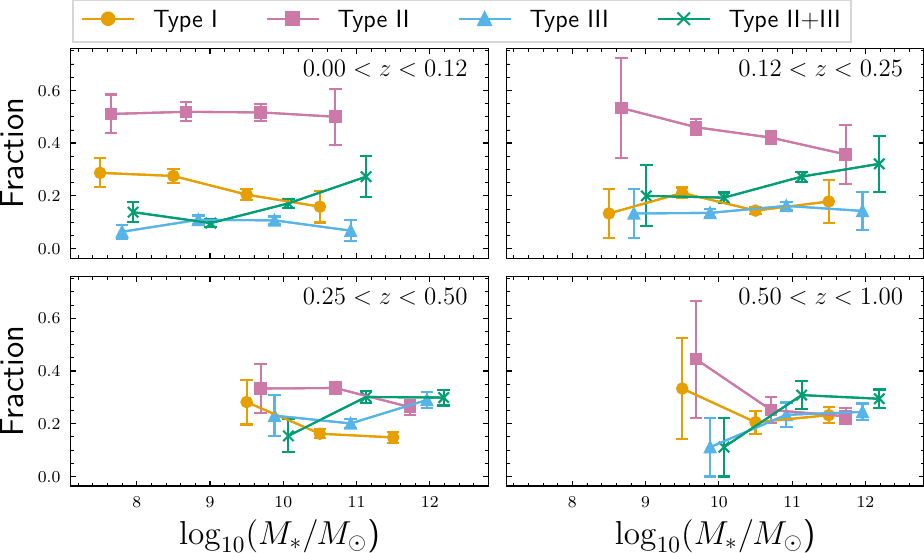}
    \caption{Fraction of disc profile types with respect to stellar mass for different redshift bins. From \textit{top} to \textit{bottom}, \textit{left} to \textit{right}, the fraction of breaks with respect to mass is shown for redshift bins of $z<0.12$, $0.12\le z<0.25$, $0.25\le z<0.50$, and $0.50\le z<1.00$, respectively. The fractions of Type~I, II, III, and II+III profiles are shown with orange circles, purple squares, blue triangles, and green cross curves, respectively.    }
    \label{fig:EUC_type_vs_mass}
\end{figure*}

The limited area coverage of the Euclid Q1 dataset introduces a selection bias in stellar mass as a function of redshift (Fig.~\ref{fig:sample}). The small area constrains the expected number of massive galaxies ($M_*>10^{10}\,M_\odot$) at lower redshifts ($z<0.06$). Since stellar mass can influence the distribution of disc break types, an effect observed in the local Universe \citep{Laine16}, we investigate the potential impact of this bias on the evolutionary trends reported in Sect.~\ref{sec:EUC_results_break_evolution}.

Figure~\ref{fig:EUC_type_vs_mass} presents the fraction of each disc break type as a function of stellar mass across four redshift bins, from the local Universe (top left panel) to higher redshift regimes (bottom right panel). If the redshift evolution observed in Fig.~\ref{fig:EUC_type_vs_z} were primarily driven by a stellar mass bias, we would expect the trends in break-type fraction versus stellar mass to remain qualitatively similar across all redshift bins if the trends found in Sect.~\ref{sec:EUC_results_break_evolution} were biased by the mass of the galaxies in our sample. Instead, we observe clearly different behaviours for Type~II and Type~III profiles. For Type~II profiles, a relatively flat trend with stellar mass is seen at low redshift ($z < 0.12$), which becomes a declining trend in the three highest redshift bins. Type~III profiles exhibit a more complex evolution: the lowest redshift bin shows a nearly flat or mildly declining fraction with stellar mass, which transitions into a flatter distribution at intermediate redshift ($0.12 < z < 0.25$), and finally into a positively correlated trend at higher redshifts ($z > 0.25$).

In contrast, Type~I and Type~II+III profiles display more consistent behaviour across redshift. Type~I profile fractions exhibit mild fluctuations, but no strong, systematic dependence on redshift or mass is observed. The Type~II+III profiles display stable trends across all redshift bins, with distributions closely resembling the global redshift evolution trend seen in Fig.~\ref{fig:EUC_type_vs_z}.

These findings suggest that the observed redshift evolution in the fractions of Type~II and Type~III discs is unlikely to be driven by stellar mass bias alone. Type~I profiles appear relatively stable, with only weak redshift-dependent variations that are also visible with mass and both parameters could drive the variations. For Type~II+III profiles, the similarity of trends across redshifts and mass implies that the evolution may reflect a combined effect, making it difficult to disentangle the contribution of each factor with the current dataset.

\begin{figure*}[t!]
    \includegraphics[width=\linewidth]{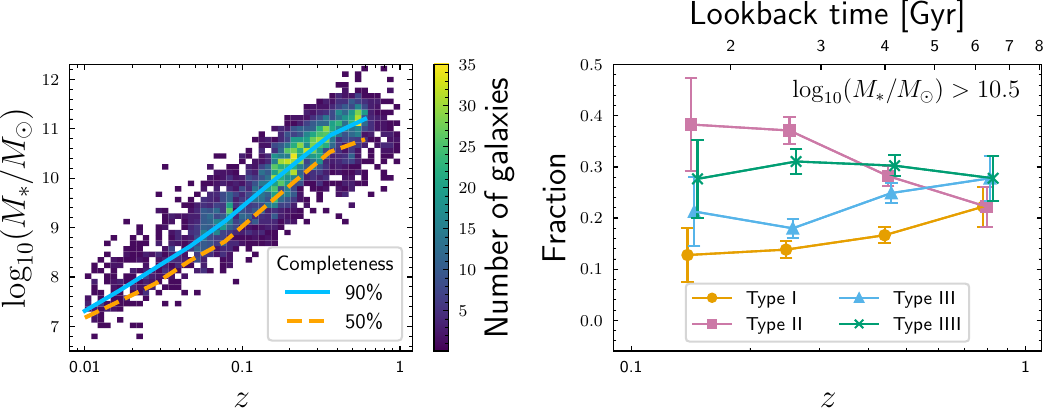}
    \caption{\emph{Left} panel: Stellar mass as a function of redshift for the galaxies in our reliable sample (Sect.~\ref{sec:EUC_Results}). The  90\% and 50\% mass completeness \citep[measured following][]{Pozzetti10} of the sample is shown in blue curves, and in orange dashed curves, respectively. \emph{Right} panel: Similar to Fig.~\ref{fig:EUC_type_vs_z} but including only galaxies with a stellar mass above $M_*>10^{10.5} M_\odot$.}
    \label{fig:EUC_completeness}
\end{figure*}

The left panel of Fig.~\ref{fig:EUC_completeness} shows the completeness, measure following \cite{Pozzetti10}, of the sample analysed in Sect.~\ref{sec:EUC_Results}. We find that the mass completeness of our sample increases rapidly with redshift. If we only select galaxies above $10^{10.5} M_\odot$, we reach a completeness of 90\% down to $z\sim0.3$. The right panel of Fig.~\ref{fig:EUC_completeness} shows the evolution of the break fraction with redshift for galaxies with a mass above  $M_*>10^{10.5} M_\odot$. The trends discussed in Sect.~\ref{sec:EUC_Results} hold, which suggests that the completeness of the sample does not drastically affect the evolutionary trends found.

\end{appendix}
\label{LastPage}
\end{document}